%% file: sample631.tex
\documentclass[onecolumn]{aastex631}

\accepted{\today}
\submitjournal{ApJ}

\usepackage{graphicx}   
\usepackage{amsmath}    
\usepackage{amssymb}    
\usepackage{CJK}        

\newcommand{\Msun}{M_{\odot}}
\newcommand{\Mbh}{M_{\rm bh}}
\newcommand{\dMdt}{{\dot{M}_{\rm in}}}
\newcommand{\Rin}{{R_{\rm in}}}
\newcommand{\Rout}{{R_{\rm out}}}
\newcommand{\rg}{{r_{g}}}
\newcommand{\sigmaSB}{{\sigma_{\rm SB}}}

\newcommand{\kB}{{k_{\rm B}}}
\newcommand{\Tvisc}{T_{\rm visc}}
\newcommand{\Tlamp}{T_{\rm lamp}}
\newcommand{\etaX}{\eta_X}
\newcommand{\etaSS}{\eta_{\rm SS}}

\newcommand{\fcol}{f_{\rm col}}
\newcommand{\Dlumi}{{D_{\rm lumi}}}
\newcommand{\LEdd}{{L_{\rm Edd}}}
\newcommand{\Lbol}{{L_{\rm bol}}}
\newcommand{\LX}{L_X}
\newcommand{\Lvisc}{{L_{\rm visc}}}
\newcommand{\Llamp}{{L_{\rm lamp}}}
\newcommand{\Tmean}{\overline{T}}
\newcommand{\Bmean}{\overline{B}_\nu}
\newcommand{\fmean}{\overline{f}_\nu}
\newcommand{\Lxmean}{\overline{L}_X}
\newcommand{\chitau}{\chi^2_\tau}
\newcommand{\chilum}{\chi^2_{\nu L_\nu}}
\newcommand{\dd}{{\rm d}}

\newcommand{\sref}[1]{Section~\ref{#1}}

\newcommand{\fref}[1]{Figure~\ref{#1}}

\newcommand{\tref}[1]{Table~\ref{#1}}

\newcommand{\eref}[1]{Equation~(\ref{#1})}
\newcommand{\esref}[1]{Equations~(\ref{#1})}

\shorttitle{RM of lamp-post and wind in thin-disk}
\shortauthors{Chan et al.}

\begin{document}
\begin{CJK*}{UTF8}{gbsn}

\title{Reverberation Mapping of Lamp-post and Wind Structures in Accretion Thin Disks}

\correspondingauthor{James~Hung-Hsu~Chan}
\email{jchan@amnh.org}

\author[0000-0001-8797-725X]{James~Hung-Hsu~Chan (詹弘旭)}
\affiliation{Department of Astrophysics, American Museum of Natural History, Central Park West and 79th Street, NY 10024-5192, USA}
\affiliation{Department of Physics and Astronomy, Lehman College of the CUNY, Bronx, NY 10468, USA}

\author[0000-0001-8723-6136]{Joshua~Fagin}
\affiliation{Department of Astrophysics, American Museum of Natural History, Central Park West and 79th Street, NY 10024-5192, USA}
\affiliation{Department of Physics and Astronomy, Lehman College of the CUNY, Bronx, NY 10468, USA}
\affiliation{The Graduate Center of the City University of New York, 365 Fifth Avenue, New York, NY 10016, USA}

\author[0009-0009-6932-6379]{Henry~Best}
\affiliation{Department of Astrophysics, American Museum of Natural History, Central Park West and 79th Street, NY 10024-5192, USA}
\affiliation{Department of Physics and Astronomy, Lehman College of the CUNY, Bronx, NY 10468, USA}
\affiliation{The Graduate Center of the City University of New York, 365 Fifth Avenue, New York, NY 10016, USA}

\author[0009-0000-4476-5003]{Matthew~J.~O'Dowd}
\affiliation{Department of Astrophysics, American Museum of Natural History, Central Park West and 79th Street, NY 10024-5192, USA}
\affiliation{Department of Physics and Astronomy, Lehman College of the CUNY, Bronx, NY 10468, USA}
\affiliation{The Graduate Center of the City University of New York, 365 Fifth Avenue, New York, NY 10016, USA}

\begin{abstract}
To address the discrepancy where disk sizes exceed those predicted by standard models, we explore two extensions to disk size estimates within the UV/optical wavelength range: disk winds and color correction. 
We provide detailed, self-consistent derivations and analytical formulas, including those based on a power-law temperature approximation, offering efficient tools for analyzing observational data.
Applying our model to four type I AGNs with intensive reverberation mapping observations, we find a shallower temperature slope ($T\propto R^{-0.66}$, compared to $R^{-3/4}$ traditionally) and a color correction factor ($f_{\rm col} \approx 1.6$), consistent with previous studies. 
We observe a positive correlation between accretion rate and color correction with black hole mass.
However, the small sample size limits our conclusions. 
The strong degeneracy between the temperature slope and accretion rate suggests that incorporating flux spectra or spectral energy distributions could improve fitting accuracy. 
Our simulation approach rapidly generates quasar light curves while accommodating various observational scenarios for reverberation mapping, making it well-suited for training machine learning algorithms.

\end{abstract}


\section{Introduction} 
\label{sec:intro}

Active Galactic Nuclei (AGNs) are astrophysical sources powered by the accretion of hot gas onto supermassive black holes (SMBHs) at the centers of galaxies. 
Gas and dust surrounding an SMBH orbit in a plane, forming what is known as an accretion disk. 
The emission from this disk is a combination of internal heat generated by viscous dissipation and external heat from the reprocessing of radiation by the corona near the SMBH.
The corona, consisting of a medium of hot electrons, Compton up-scatters radiation, which is subsequently emitted as X-rays \citep[e.g.][]{Lightman&White88}.
In the standard theory of AGN, the central accretion disk is typically modeled as optically thick and geometrically thin
\citep{Lynden-Bell69,Pringle&Rees72,Shakura&Sunyaev73,Novikov&Thorne73,Friedjung85}, emitting a blackbody continuum that exhibits variability.  
While the physical origin of this variability remains unclear, several studies indicate that ultraviolet (UV) to optical variability may be driven by preceding X-ray variations \citep[e.g.][]{CackettEtal07,McHardyEtal14}. 

The reprocessing of the driving X-ray variability into the UV/optical emitting regions of the accretion disk is known as continuum reverberation mapping. 
Reverberation mapping, in general, encompasses several types, each probing different regions of AGN \citep[see the review by][]{CackettEtal21}. 
For instance, continuum reverberation mapping focuses on the accretion disk, emission-line reverberation mapping probes the structure and kinematics of the broad-line and narrow-line regions (BLR and NLR), and dust reverberation mapping explores the torus surrounding the AGN.
This phenomenon was first observed through the time lag between the continuum emission and the H$\alpha$ emission lines \citep{Cherepashchuk&Lyutyi73}. 
The theoretical framework for this technique was later established by \cite{Blandford&Mckee82}, and it has since been applied to observations to estimate the size of BLR \citep{Gaskell&Sparke86}. 
Assuming that broad-line emission is triggered by central emission, the lag represents the light-travel time from the central illuminating source to the BLR. 
This allows for calculating the BLR's size as $R_{\rm BLR}=c\ \tau_{\rm lag}$. 
Similarly, this method can be applied to determine the size of the accretion disk, as the continuum emission across the disk is also driven by the central source. 
The time lags measured through continuum reverberation mapping of UV/optical light curves provide insights into the relative size scales of the emitting regions. 
These measurements are closely linked to the properties of the accretion disk and the black hole \citep[e.g.][]{FausnaughEtal17}.
In the remainder of this work, we refer to reverberation mapping specifically as continuum reverberation mapping of the accretion disk.

To understand the growth and evolution of SMBHs in AGNs, it is crucial to study their accretion disk structures. 
Studies on quasar variability has shown that accretion disk sizes are larger than those predicted by the thin-disk model, by a factor of about 2--4 \citep{FausnaughEtal16,JiangEtal17, MuddEtal18,YuEtal20,GuoEtal22,JhaEtal22}. 
This observation is also supported by accretion disk size measurements obtained through gravitational microlensing~\citep{MorganEtal10,BlackburneEtal11,MunozEtal16,MorganEtal18,Cornachione&Morgan20}.
Some studies suggest that this size discrepancy may be attributed to factors such as an underestimation of black hole mass, which could be larger by a factor of 3.3 when the geometric scaling factor is 6 times greater \citep{PozoNunezEtal19}, or the negligence of reddening, which can underestimate the disk size by a factor of 2.4 \citep{GaskellEtal23}.
In addition, the optical flux measured from the source is significantly lower than that predicted by a standard geometrically thin disk emitting at the observed wavelength. 
To explain this discrepancy, we explore several extensions of the models for UV/optical reverberation mapping, which have also been applied in microlensing analysis~\citep{ZdziarskiEtal22}.

We first consider mass loss due to winds from the disk surface, which can significantly alter the predicted mass accretion rate between disks with and without wind, thereby reconciling theoretical predictions with observations \citep{LiEtal19,SunEtal19}.
A second effect involves local color corrections to the disk's blackbody emission, with $\fcol\geq1$. 
Several studies have explored the concept of accretion disks sustained by magnetic pressure, which are substantially hotter than traditional thin disk models \citep{BlaesEtal06,Begelman&Pringle07,SalvesenEtal16}.
While some studies have found $\fcol\approx1$ in optical wavelengths~\citep{HubenyEtal01}, this assumes no dissipation in the disk's surface layers. 
Relaxing this assumption to allow for moderate dissipation in the surface layers can potentially resolve the observed discrepancy~\citep[e.g.][]{RossEtal92,KammounEtal21a}.
In this work, we study the effects of these phenomena within the frameworks of the standard accretion disk model and the lamp-post model, with a particular focus on their implications for UV/optical reverberation mapping.

This paper is organized as follows. 
In \sref{sec:theory}, we derive the physical quantities for the thin-disk, lamp-post, and wind models. 
The results of the observed time-lag spectra, analyzed using our model, are presented in \sref{sec:data}. 
We discuss model parameters in \sref{sec:discussion} and conclude our findings in \sref{sec:conclusion}.
When required, a flat cosmology is used with $H_0=70~{\rm km~s^{-1}~Mpc^{-1}}$, $\Omega_m=0.3$ and $\Omega_\Lambda=0.7$.

\begin{figure*}
\centering
\includegraphics[width=\textwidth]{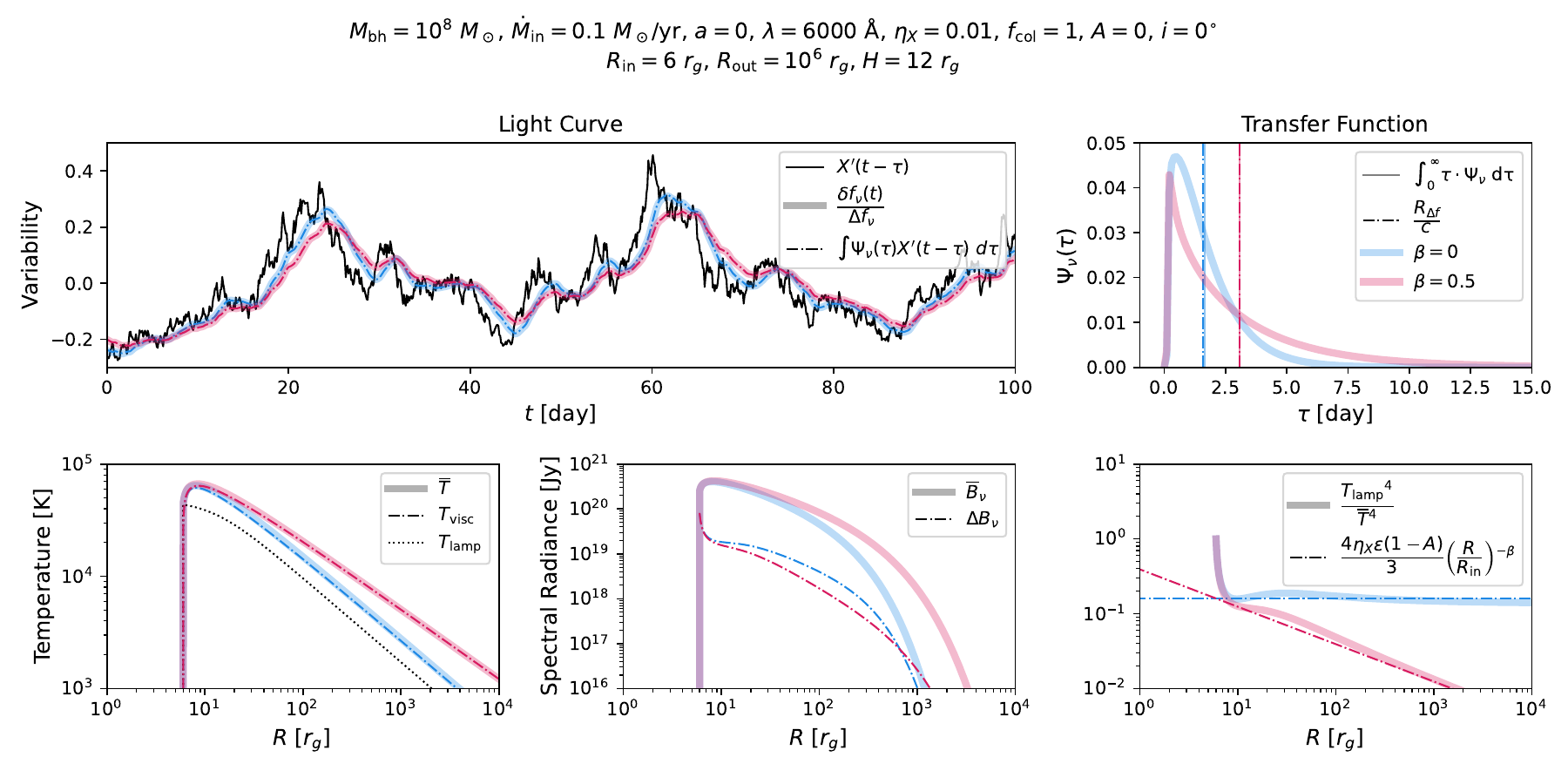}
\caption{
\textit{Top-left}: Simulated light curves. 
The black curve represents X-ray variability, the solid color curves exhibit the light curves from perturbed temperature using $B_\nu(T,t)$, and dash-dotted color curves show the light curves from convolution of the transfer functions. 
\textit{Top-right}: Transfer functions.
The solid vertical lines indicate the average time lags $\tau_\lambda$ using \eref{eqn:time_lag} and the dash-dotted vertical lines label the estimation using power law approximation of \eref{eqn:time_lag_pl}.
\textit{Bottom-left}: Effective temperature $\Tmean$ (solid line) profiles as the sum of viscous heating $\Tvisc$ (dash-dotted line) and lamp-post irradiation $\Tlamp$ (dotted line).
Note that the slope of $\Tvisc$ follows the power index of $-\frac{3-\beta}{4}$.
\textit{Bottom-middle}: Spectral radiance calculated using \esref{eqn:B_nu} and (\ref{eqn:dB_nu}). 
\textit{Bottom-right}: Ratio of $\Tlamp$ and $\Tmean$, which is crucial for computing $\Delta B_\nu$.
The dash-dotted lines indicate the approximation given by \eref{eqn:T4_apprx}, which is used in the calculation of $R_{\Delta f}$ in \eref{eqn:time_lag_pl}.
The red and blue colors represent two power indices for the accretion rate $\dot{M}(R)=\dMdt\left(R/\Rin\right)^\beta$, with $\beta=0$ and $\beta=0.5$, respectively.
The accretion disk parameters are listed in the title. 
The Eddington ratios are $\Lbol/\LEdd=0.0375$ and $0.1121$ for $\beta=0$ and $0.5$, respectively.
The Eddington ratio of the X-ray lamp post can be obtained as $\Lxmean/\LEdd=0.0045$.
}
\label{fig:output}
\end{figure*}
\section{Theory}
\label{sec:theory}
The structure of an AGN can be postulated as a geometrically thin, optically thick accretion disk irradiated by external sources.
This configuration predicts specific dependencies of observed time lags and fluxes on wavelength $\lambda$ (or frequency $\nu=c/\lambda$), expressed as the time-lag spectrum $\tau_\lambda$ and the flux spectrum $f_\nu$. 
By measuring these spectra, we can infer the physical structure of the accretion disk.
This section presents a comprehensive calculation of light curves for an accretion disk model irradiated by a ``lamp post'' geometry (see \sref{sec:reverberation}).
We build upon previous works on irradiated disk light curves \citep[e.g.][]{CackettEtal07,StarkeyEtal16,KovacevicEtal22,PozoNunezEtal23} by introducing an approach that incorporates a power-law dependence for the accretion rate and explores the influence of the disk structure.  
Furthermore, we provide analytical formulas derived using a power-law approximation of the temperature profile, allowing for more efficient application to observational data (see \sref{sec:temperature_powerlaw}).

\subsection{Thermal Reverberation in a Thin-Disk with Wind Systems}
\label{sec:reverberation}
Given a central SMBH of AGN with mass $\Mbh$, the inner limit of the accretion disk is $\Rin = \alpha~\rg$, where $\alpha$ is determined by the spin $a$ within the Kerr metric \citep{Kerr63,Bardeen72} and the gravitational radius is defined as $\rg=G\Mbh/c^2$. 
This inner limit $\Rin$ is also known as innermost stable circular orbit (ISCO) radius. 
For example, $\alpha = 6$ for a non-spinning black hole ($a = 0$), while $\alpha = 1$ for a maximally spinning black hole ($a = 1$).
To account for mass loss in the wind, we adopt a power-law dependence for the mass accretion rate of the disk \citep{Blandford&Begelman99}:
\begin{equation}
    \dot{M}(R)=\dMdt\left(\frac{R}{\Rin}\right)^\beta,
    \label{eqn:dMdt}
\end{equation}
where $\dMdt$ is the mass accretion rate at $\Rin$ and $\beta$ is a constant power-law index ($0\leq\beta<1$, ensuring that the mass accretion rate decreases while the released energy increases with accretion).
This approach represents a self-similar advection-dominated inflow–outflow solution (ADIOS) for advection-dominated accretion flows (ADAF) with winds, which has been further supported by numerical simulations of accretion disks \citep{Xie&Yuan08,Li&Cao09}.
For a standard thin-disk model \citep[SS disk;][]{Shakura&Sunyaev73} irradiated by a lamp-post geometry X-ray corona with luminosity $\LX$ located above the SMBH at a height of $H=\epsilon~\rg$, the disk temperature with albedo $A$ can be expressed as:
\begin{equation}
    T^4(R,\theta,t)=\frac{3G\Mbh \dot{M}}{8\pi\sigmaSB R^3}\left(1-\sqrt{\frac{\Rin}{R}}\right)
    +\frac{2\cdot(1-A)\cdot \LX(t-\tau)}{4\pi\sigmaSB}\frac{H}{\left(R^2+H^2\right)^{3/2}},
\label{eqn:temp}
\end{equation}
where the Stefan-Boltzmann constant is given by $\sigmaSB=\dfrac{2\pi^5\kB^4}{15h^3c^2}$ with the Boltzmann constant $\kB$.
The X-ray corona is a variable source $\LX(t)=\Lxmean\cdot X(t)=\etaX\dMdt c^2\cdot X(t)$, where $\etaX$ is radiative efficiency in X-ray, which heats up the disk with a time lag $\tau$ at different positions:
\begin{equation}
    c\cdot\tau(R,\theta) = \sqrt{R^2+H^2}+H\cos i-R\cos\theta\sin i,
    \label{eqn:tau}
\end{equation}
where $R$ and $\theta$ are coordinates of disk with inclination angle $i$.
The variability function over time $t$ is denoted as $X(t)$, which is reasonably well described by a damped random walk \citep{KellyEtal09,KozlowskiEtal10,ZuEtal13}.
We impose the convention that the mean and standard deviation of $X(t)$ are both unity in this work, acknowledging that this selection may vary across different studies.
The factor of $2$ in the second term on the right hand side of \eref{eqn:temp} is considered as two sides of lamp post \citep[e.g.][]{DovciakEtal22,Liu&Qiao22}.
We note that the lamp-post model may not be limited to the X-ray corona; it can also encompass other sources of variability. 
However, we choose this approach for simplicity (see the discussion in \sref{sec:discussion}).

To account for small temperature perturbations, we define the effective temperature as a combination of viscous heating $\Tvisc$ and lamp-post irradiation $\Tlamp$, expressed as
\begin{equation}
\begin{split}
    &{\Tmean}^4(R)={\Tvisc}^4(R)+{\Tlamp}^4(R)\\
    &=\left(\frac{3G\Mbh\dMdt}{8\pi\sigmaSB\Rin^3}\right)
    \left[\left(\frac{\Rin}{R}\right)^{3-\beta}\left(1-\sqrt{\frac{\Rin}{R}}\right)+\frac{4\etaX\epsilon(1-A)\Rin^3}{3\left(R^2+H^2\right)^{3/2}}\right].\\
\end{split}
\label{eqn:teff}
\end{equation} 
Subsequently, the temperature of \eref{eqn:temp} can be described using the first-order Taylor expansion as 
\begin{equation}
T(R,t) \approx \Tmean\left[1+\frac{{\Tlamp}^4}{4{\Tmean}^4}\cdot\left(X(t-\tau)-1\right)\right]
= \Tmean+\Delta T\cdot X'(t-\tau),
\label{eqn:temp_apprx}
\end{equation}
where the amplitude of temperature fluctuation is approximated as
$\Delta T={\Tlamp}^4/4{\Tmean}^3$ and the variability function of temperature is defined as $X'(t)=X(t)-1$.
Considering blackbody radiation with color correction $\fcol$ at frequency $\nu$ \citep[e.g][]{Shimura&Takahara95}, the expression is given by:
\begin{equation}
    B_\nu(T,t)
    =\frac{2h\nu^3}{{\fcol}^4 c^2}\frac{1}{e^{\frac{h\nu}{\fcol\kB T}}-1}
    \approx \Bmean(R)+\Delta B_\nu(R)\cdot X'(t-\tau),
\end{equation}
where the time independent term is defined as:
\begin{equation}
    \Bmean(R)=\frac{2h\nu^3}{{\fcol}^4 c^2}\frac{1}{e^{\xi}-1},
\label{eqn:B_nu}
\end{equation}
and the amplitude of the time-dependent component is determined by:
\begin{equation}
    \Delta B_\nu(R)
    =\frac{\partial B_\nu}{\partial T}\Big|_{\Tmean}\cdot\Delta T
    =\frac{h\nu^3}{2{\fcol}^4 c^2}\frac{\xi e^{\xi}}{\left(e^{\xi}-1\right)^2}\frac{{\Tlamp}^4}{{\Tmean}^4}.
\label{eqn:dB_nu}
\end{equation}
In this context, we adopt the definition:
\begin{equation}
    \xi=\frac{h\nu}{\fcol\kB \Tmean}.
    \label{eqn:xi}
\end{equation}
Hence, the variation in flux $\delta f_\nu(t)$ can be simplified as a convolution:
\begin{equation}
    \delta f_\nu(t) = f_\nu(t)-\fmean \approx \Delta f_\nu\cdot \int_0^\infty \Psi_\nu(\tau|\lambda)\ X'(t-\tau)\ \dd\tau,
    \label{eqn:dflux_conv}
\end{equation}
where $f_\nu(t)$ represents the total flux, $\fmean$ denotes the mean flux, and $\Delta f_\nu$ stands for the amplitude of flux variation:
\begin{align}
    f_\nu(t) &= \int_{\Omega} B_\nu(R,t)\ \dd\Omega, \label{eqn:flux(t)}\\
    \fmean  &= \int_{\Omega} \Bmean(R)\ \dd\Omega, \label{eqn:flux}\\
    \Delta f_\nu &= \int_{\Omega} \Delta B_\nu(R)\ \dd\Omega \label{eqn:dflux},
\end{align}    
where the solid angle $\dd\Omega = R\ \dd\theta\ \dd R\ \cos i / \Dlumi^2$ with the luminosity distance $\Dlumi$, and the transfer function at wavelength $\lambda$ (or frequency $\nu$) is given by
\begin{equation}
    \Psi_\nu(\tau|\lambda)=\frac{1}{\Delta f_\nu}\int_{\Omega'} \Delta B_\nu(R')\ \delta(\tau-\tau')\ \dd \Omega'.
\end{equation}
The transfer function describes how the disk responds to changes at a specific wavelength and time lag, normalized to unity: $\int_0^\infty \Psi_\nu(\tau|\lambda)\ \dd\tau=1$.
Consequently, the average time lag at a given wavelength can be calculated as:
\begin{equation}
    \tau_\lambda
    = \int_0^\infty \tau\cdot\Psi_\nu(\tau|\lambda)\ \dd\tau
    =\frac{\int R\cdot\Delta B_\nu(R)\ \dd\Omega}{c\cdot\Delta f_\nu}.
\label{eqn:time_lag}
\end{equation}
We emphasize that the variability derived using the transfer function is valid when temperature fluctuations are small in \eref{eqn:temp_apprx} (i.e. $\Tlamp<\Tvisc$). 
In \fref{fig:output}, we illustrate light curves from both full simulation and using the transfer function, demonstrating the effectiveness of the convolution approach.

\begin{figure*}[t]
\centering
\includegraphics[width=\textwidth]{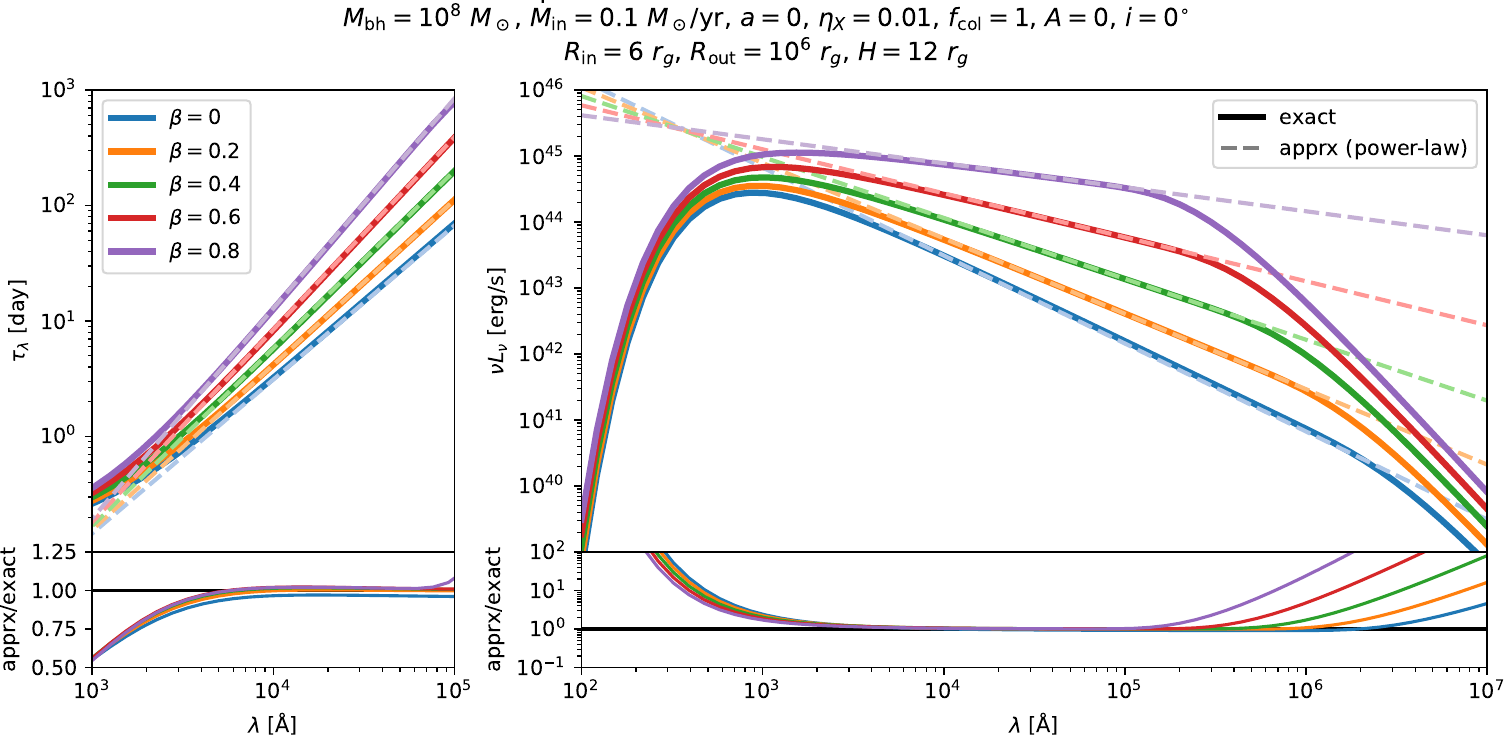}
\caption{
Spectra of average time lag (\textit{left}) and luminosity (\textit{right}). 
The solid lines are obtained from numerical calculations (exact), while the dashed lines are derived from the power law approximation (apprx), yielding $\tau_\lambda\propto \lambda^{\frac{4}{3-\beta}}$ and $\nu L_\nu \propto\lambda^{-\frac{4-4\beta}{3-\beta}}$. 
The bottom panels illustrate the comparison between two methods. 
The discrepancy at wavelengths $\lesssim3000~\textrm{\AA}$ is attributed to the ISCO size $\Rin$ and the corona height $H$, while that at wavelengths $\gtrsim10^5~\textrm{\AA}$ is due to the outer radius $\Rout$ (i.e. the truncation of the disk). 
Both methods agree well when $\Rin, H \ll R \ll \Rout$. 
This analysis demonstrates that for typical AGNs, the power-law approximation is effective within the wavelength range of approximately 3000 to $10^5~\textrm{\AA}$~(UV-optical).
The accretion disk parameters are listed in the title, same as \fref{fig:output}.
}
\label{fig:lamb}
\end{figure*}
Finally, we present the expression for the bolometric luminosity of the accretion disk at a luminosity distance $\Dlumi$ with an outer radius $\Rout$:
\begin{equation}
\begin{split}
    \Lbol
    &=\frac{2\pi \Dlumi^2}{\cos i}\int_0^\infty \fmean\ \dd\nu
    =2\sigmaSB\int_0^{2\pi}\int_\Rin^\Rout \Tmean^4(R) \ R\ \dd R\ \dd\theta\\
    &=2\sigmaSB\int_0^{2\pi}\int_\Rin^\Rout \left({\Tvisc}^4+{\Tlamp}^4\right) \ R\ \dd R\ \dd\theta 
    =\Lvisc+\Llamp,\\
\end{split}
\end{equation}
where the contributions from viscous heating $\Lvisc$ and lamp-post irradiation $\Llamp$ are given by:
\begin{equation}
    \Lvisc=\frac{G\Mbh\dMdt}{2\Rin}
    \left[\frac{3}{\left(1-\beta\right)\left(3-2\beta\right)}-\frac{3}{1-\beta}\left(\dfrac{\Rin}{\Rout}\right)^{1-\beta}+\frac{6}{3-2\beta}\left(\dfrac{\Rin}{\Rout}\right)^{\frac{3-2\beta}{2}}\right],
    \label{eqn:lumi_visc}
\end{equation}
\begin{equation}    
    \Llamp=4\etaX\epsilon(1-A)\frac{G\Mbh\dMdt}{2\Rin}
    \left[ \left(1+\frac{H^2}{\Rin^2}\right)^{-\frac{1}{2}}-\left(\frac{\Rout^2}{\Rin^2}+\frac{H^2}{\Rin^2}\right)^{-\frac{1}{2}}\right].
    \label{eqn:lumi_lamp}
\end{equation}
For the scenario where there is no wind (i.e. $\beta=0$, $\dMdt=\dot{M}$), the outer radius greatly exceeds the inner radius ($\Rout\gg\Rin$), and irradiation effects are negligible (e.g. $\etaX,\epsilon\ll1$), the bolometric luminosity can be approximated as:
\begin{equation}
    \Lbol=\frac{G\Mbh\dMdt}{2\Rin}=\frac{\dot{M} c^2}{2\alpha} = \etaSS \dot{M} c^2,
\end{equation}
where the radiative efficiency for an SS disk is denoted as $\etaSS \equiv 1/2\alpha$.

\subsection{Power-Law Temperature Approximation}
\label{sec:temperature_powerlaw}
In the UV/optical regime, the effects of the lamp post and ISCO on the disk temperature are typically negligible, i.e. $H, \Rin \ll R$.
The effective temperature of \eref{eqn:teff} can then be approximated by a simple power law \citep[e.g.][]{Montesinos12}:
\begin{equation}
\Tmean^4\approx\left(\dfrac{3G\Mbh\dMdt}{8\pi\sigmaSB\Rin^3}\right)
    \left(\dfrac{\Rin}{R}\right)^{3-\beta}.    
\end{equation}
Assuming $\Rout\gg R$, the mean flux $\fmean$ and the amplitude of variation $\Delta f_\nu$ can be calculated using \esref{eqn:flux} and (\ref{eqn:dflux}) as follows:
\begin{equation}
    \fmean \approx
    \frac{4\pi \cos i }{\Dlumi^2}
    \left[\frac{4}{3-\beta}\zeta(\frac{8}{3-\beta})\Gamma(\frac{8}{3-\beta})\right]
    \left[\left(\frac{45G\Mbh\dMdt}{16\pi^6c}\right)^2\left(\frac{\alpha G\Mbh}{c^2}\right)^{-2\beta}\left(\frac{c}{\nu}\right)^{2\beta}\left(\frac{h\nu}{{\fcol}^4}\right)^{1-\beta}\right]^{\frac{1}{3-\beta}},
    \label{eqn:flux_pl}
\end{equation}
\begin{equation}
    \Delta f_\nu \approx
    \frac{4\pi \cos i }{\Dlumi^2}\frac{\etaX\epsilon(1-A)}{3} 
    \left[\frac{4}{3-\beta}\zeta(\frac{8-4\beta}{3-\beta})\Gamma(\frac{11-5\beta}{3-\beta})\right]
    \left[\left(\frac{45 G\Mbh\dMdt}{16\pi^6 c}\right)^{2-\beta}
\left(\frac{\alpha G\Mbh}{c^2}\right)^{\beta}\left(\frac{c}{\nu}\right)^{-\beta}
\left(\frac{h\nu}{{\fcol}^4}\right)
\right]^{\frac{1}{3-\beta}}.\\
\label{eqn:dflux_pl}
\end{equation}
In the previous calculations, we utilize the following equations involving the Riemann zeta function $\zeta$ and the gamma function $\Gamma$:
\begin{equation}
    \int_0^\infty\frac{x^b}{e^x-1}\ \dd x= \zeta(b+1)\ \Gamma(b+1) \quad \text{and} \quad
    \int_0^\infty\frac{x^b e^x}{\left(e^x-1\right)^2}\ \dd x= \zeta(b)\ \Gamma(b+1).
\end{equation}
Note that for $\Delta f_\nu$ we introduce an additional approximation for variational radiation $\Delta B_\nu$ in \eref{eqn:dB_nu}, given by: 
\begin{equation}
    \frac{{\Tlamp}^4}{{\Tmean}^4}
\approx\frac{4\etaX\epsilon(1-A)}{3}\left(\frac{R}{\Rin}\right)^{-\beta}.
\label{eqn:T4_apprx}
\end{equation}
This approximation is valid under the conditions $\Rin, H \ll R$, and $\etaX\epsilon(1-A) \ll 1$, consistent with our power-law temperature assumption.
We can also convert flux to luminosity using the relation $\nu L_\nu = \lambda L_\lambda=\dfrac{2\pi \Dlumi^2}{\cos i} \nu \fmean$. 

\begin{figure*}[t]
\centering
\includegraphics[width=\textwidth]{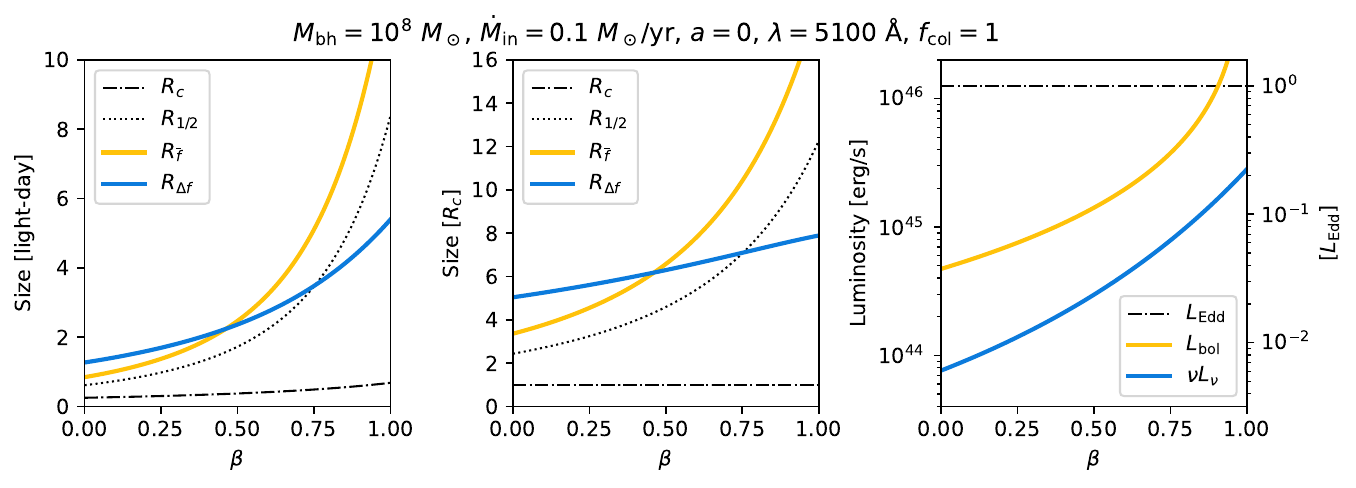}
\caption{
Disk size (\textit{left, middle}) and luminosity (\textit{right}) against accretion rate slope.
The values are obtained using power law approximation, where $\Rin,H\ll R\ll \Rout$.
The characteristic radius $R_c$, half-light $R_{1/2}$, flux-weighted radius $R_{\overline{f}}$, and flux-variation-weighted radius $R_{\Delta f}$ are calculated following \esref{eqn:R_c}, (\ref{eqn:R_hl}), (\ref{eqn:R_fl}), and (\ref{eqn:R_df}), respectively.
The left and middle panels differ only due to the use of varying units.
Note that when $\beta=0$, $R_{1/2}/R_c=2.44$, $R_{\overline{f}}/R_c=3.36$, and $R_{\Delta f}/R_c=5.04$, respectively.
The monochromatic luminosity $\nu L_\nu$ is obtained following \eref{eqn:flux_pl}, using the conversion $\left(2\pi \Dlumi^2/\cos i \right)\nu \fmean$.
The bolometric luminosity $\Lbol$ ($=\Lvisc$) is calculated using \eref{eqn:lumi_visc} under the condition $\Rout\gg\Rin$.
The right panel displays units on both the left and right sides.
The Eddington luminosity is denoted as $\LEdd$.
}
\label{fig:beta}
\end{figure*}
Next, we derive the various physical sizes using the relation in \eref{eqn:xi}: 
\begin{equation}
    \xi = \frac{h\nu}{\fcol\kB \Tmean} = \left(\frac{R}{R_c}\right)^{\frac{3-\beta}{4}},
\end{equation}
where $R_c$ is defined as the characteristic radius at which the disk temperature matches the frequency (or wavelength): $h\nu=\fcol\kB\Tmean$.
This radius $R_c$ can then be expressed as:
\begin{equation}
\begin{split}
    R_c &= 
\left(\frac{45G\Mbh\dMdt}{16\pi^6hc^2}\right)^{\frac{1}{3-\beta}}
\left(\frac{\alpha G\Mbh}{c^2}\right)^{-\frac{\beta}{3-\beta}}\left(\fcol\cdot\lambda\right)^{\frac{4}{3-\beta}}.\\
\end{split}
\label{eqn:R_c}
\end{equation}
For the case of $\beta=0$, it has been widely used in several works in the literature \citep[e.g.][]{MorganEtal10}.
It is straightforward to obtain the half-light radius $R_{1/2}$, the flux-weighted radius $R_{\overline{f}}$, and the flux-variation-weighted radius $R_{\Delta f}$ as follows:
\begin{equation}
    \int_0^{2\pi}\int_0^{R_{1/2}} \Bmean(R)\ \dd\Omega=\frac{\fmean}{2},
    \label{eqn:R_hl}
\end{equation}
\begin{equation}
    R_{\overline{f}}
    =\frac{\int R\cdot \Bmean(R)\ \dd\Omega}{\int \Bmean(R)\ \dd\Omega}
    =R_c \cdot \frac{\zeta(\frac{12}{3-\beta})\Gamma(\frac{12}{3-\beta})}{\zeta(\frac{8}{3-\beta})\Gamma(\frac{8}{3-\beta})},
    \label{eqn:R_fl}
\end{equation}
\begin{equation}
    R_{\Delta f}
    =\frac{\int R\cdot \Delta B_\nu(R)\ \dd\Omega}{\int\Delta B_\nu(R)\ \dd\Omega}
    =R_c \cdot \frac{\zeta(\frac{12-4\beta}{3-\beta})\Gamma(\frac{15-5\beta}{3-\beta})}{\zeta(\frac{8-4\beta}{3-\beta})\Gamma(\frac{11-5\beta}{3-\beta})}.
    \label{eqn:R_df}
\end{equation}
Here, we apply the approximation of \eref{eqn:T4_apprx} for $R_{\Delta f}$ as well.
We notice that a similar derivation is presented in \cite{GuoEtal22}; however, the method employed in this work offers a more self-consistent approach.
For the case where $\beta=0$, $R_{1/2}=2.44~R_c$, $R_{\overline{f}}=3.36~R_c$, and $R_{\Delta f}=5.04~R_c$, which has been extensively used in several studies \citep[e.g.][]{MuddEtal18,Tie&Kochanek18,YuEtal20}.
We emphasize that in reverberation mapping applications, it is important to carefully consider the selection of the flux-variation-weighted radius $R_{\Delta f}$ for time lag measurement, ensuring its alignment with the mean of the transfer function:
\begin{equation}
    \tau_\lambda=\frac{R_{\Delta f}}{c} = \frac{R_c}{c}\frac{\zeta(\frac{12-4\beta}{3-\beta})\Gamma(\frac{15-5\beta}{3-\beta})}{\zeta(\frac{8-4\beta}{3-\beta})\Gamma(\frac{11-5\beta}{3-\beta})}.
    \label{eqn:time_lag_pl}
\end{equation}
This relationship is illustrated by the dash-dotted vertical line in the top-right panel of \fref{fig:output}, showing good agreement with the solid vertical line from \eref{eqn:time_lag}.
However, it is worth noting that most traditional curve-shifting techniques can underestimate the mean time lag by up to $\approx30\%$ \citep{ChanEtal19}.
From the power index $\beta$ defined for the accretion rate,
we derive its implications for the temperature slope $T\propto R^{-\frac{3-\beta}{4}}$, the slope of the time-lag spectrum $\tau_\lambda\propto R_c \propto \lambda^{\frac{4}{3-\beta}}$, and the luminosity spectrum $\nu L_\nu \propto\lambda^{-\frac{4-4\beta}{3-\beta}}$.
Although the examples shown in this work are limited to the range of $0\leq\beta<1$, we note that the equations of this section can be extended to any slope.
Particularly for $\beta>1$, it implies a more significant impact of the wind on the disk, as the wind could be more effective at removing material from the disk at larger radii. 
Such strong winds, however, might be less likely in practice.
In addition, some equations would require modification for broader applicability (e.g. \eref{eqn:lumi_visc}\footnote{\begin{align*}
\Lvisc=\frac{G\Mbh\dMdt}{2\Rin}\times
\begin{cases}
&    \left[3\ln\left(\Rout/\Rin\right)+6\sqrt{\Rin/\Rout}-6\right],\ \ \beta=1\\
&    \left[3\ln\left(\Rin/\Rout\right)+6\sqrt{\Rout/\Rin}-6\right],\ \ \beta=3/2\\
\end{cases}
\end{align*} 
\center{($\Lvisc$ can diverge when $\beta\geq1$ and $\Rout\to\infty$.)}
}).
It is crucial to assess the validity of the chosen parameters, particularly $\Rin$, $\Rout$, and $H$, as they can introduce bias in the power-law approximation (see the discussion in \sref{sec:discussion}).

In \fref{fig:lamb}, we demonstrate the validation of power-law approximation on time-lag and luminosity spectra.
For the UV-optical range, the power-law approximation captures the trend well.
At short wavelengths ($\lambda\lesssim 1000~\textrm{\AA}$), it tends to underestimate time lags and overestimate luminosities, partially due to the ISCO size, although this effect can be mitigated in high spin AGNs (e.g. $a=1$ and $\alpha=1$).
The turnover in luminosity (and also time lag) at longer wavelengths ($\lambda\gtrsim 10^5~\textrm{\AA}$) is influenced by the outer radius $\Rout$.
To approach the power-law approximation more closely, increasing $\Rout$ could alleviate the issue, although the exact value may not be determinable.
In \fref{fig:beta}, we illustrate how $\beta$ enhances size and luminosity.

\section{Data fitting}
\label{sec:data}
\begin{table*}[b]
  \caption{
  AGN sample properties.
  }
  \input{table/case.tex}
  \tablecomments{
  $\Mbh$ and $z$ are obtained from The AGN Black Hole Mass Database\footnote{\url{http://www.astro.gsu.edu/AGNmass/}} \citep{Bentz&Katz15}, except for NCG 5548 where $\Mbh$ is from \cite{HorneEtal21}.
  $\Dlumi$ represents the luminosity distance at redshift $z$.
  $\nu L_\nu(5100\textrm{\AA})$ are taken from \cite{Netzer22}.
  $\lambda_{\rm ref}$ is the reference wavelength in the observed frame used for measuring time lag spectra.
  }
  \label{tab:case}
\end{table*}
To evaluate the practical efficacy of the power-law approximation, we fit the time lags of AGNs with well-sampled light curves. 
Our sample comprises (type I) AGNs with known time-lag spectra obtained from intensive, multi-wavelength campaigns: NGC~4593 \citep{CackettEtal18}, NGC~5548 \citep{FausnaughEtal16}, Fairall~9 \citep{HernandezSantistebanEtal20}, and Mrk~817 \citep{KaraEtal21}.
These four AGNs have also been used to demonstrated the work in \cite{KammounEtal23}.
Furthermore, we utilize luminosity measurements at $\lambda=5100~\textrm{\AA}$ in the rest frame from \cite{Netzer22} as prior information, aiding in the constraint of the power index $\beta$. 
Further details of this sample are listed in \tref{tab:case}. 
We note that for all time lag measurements, we adopt the centroid values of the ICCF as reported in the corresponding papers.
\begin{figure}
\centering
\includegraphics[width=\textwidth]{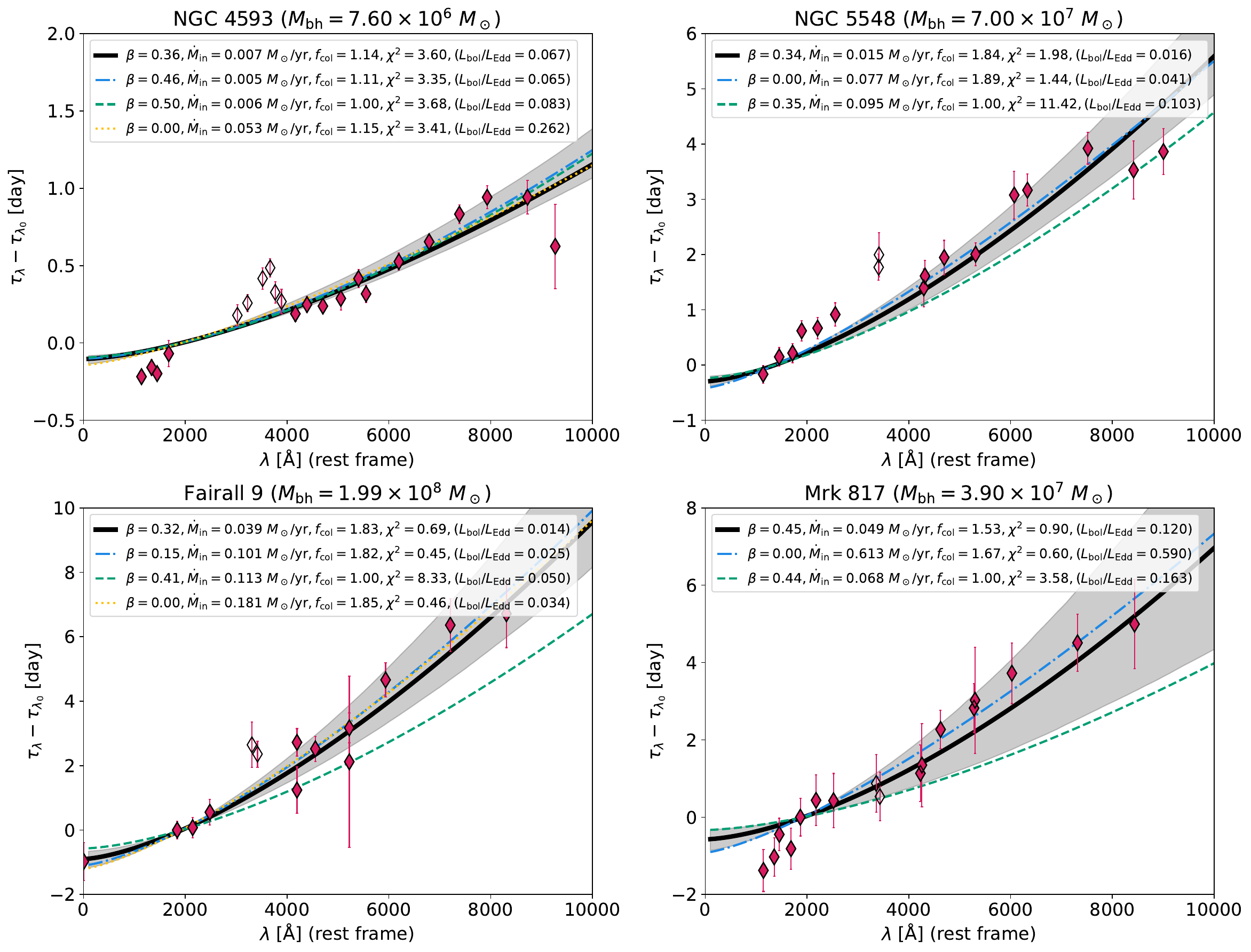}
\caption{
Fitting results for the AGN sample.
The total chi-square is calculated as $\chi^2=\chitau+\chilum$ using \esref{eqn:chitau} and (\ref{eqn:chilum}).
Time lags and luminosities, shifted to the rest frame, are determined using a power-law temperature approximation (see \sref{sec:temperature_powerlaw}).
The spin is fixed as free, i.e. $a=0$ and $\alpha=6$; however, this choice can degenerate with $\dMdt$ (i.e. ${R_c}^{3-\beta}\propto {\nu L_\nu}^{\frac{3-\beta}{2}} \propto\dMdt\alpha^{-\beta}$). 
Solid black lines represent the MCMC result with $1\sigma$ uncertainty in grey shades. 
Dash-dotted blue lines depict the Powell best-fit, while dashed green lines illustrate the best-fit without color correction ($\fcol=1$), demonstrating the importance of including this parameter, and dotted yellow lines (if needed) show the best-fit without wind effects ($\beta=0$). 
Open diamonds label time-lag measurements excluded from our fitting process due to potential influence from the significantly larger Broad Line Region (BLR) emissions.
}
\label{fig:case}
\end{figure}
\begin{figure}
    \centering
    \includegraphics[width=0.48\textwidth]{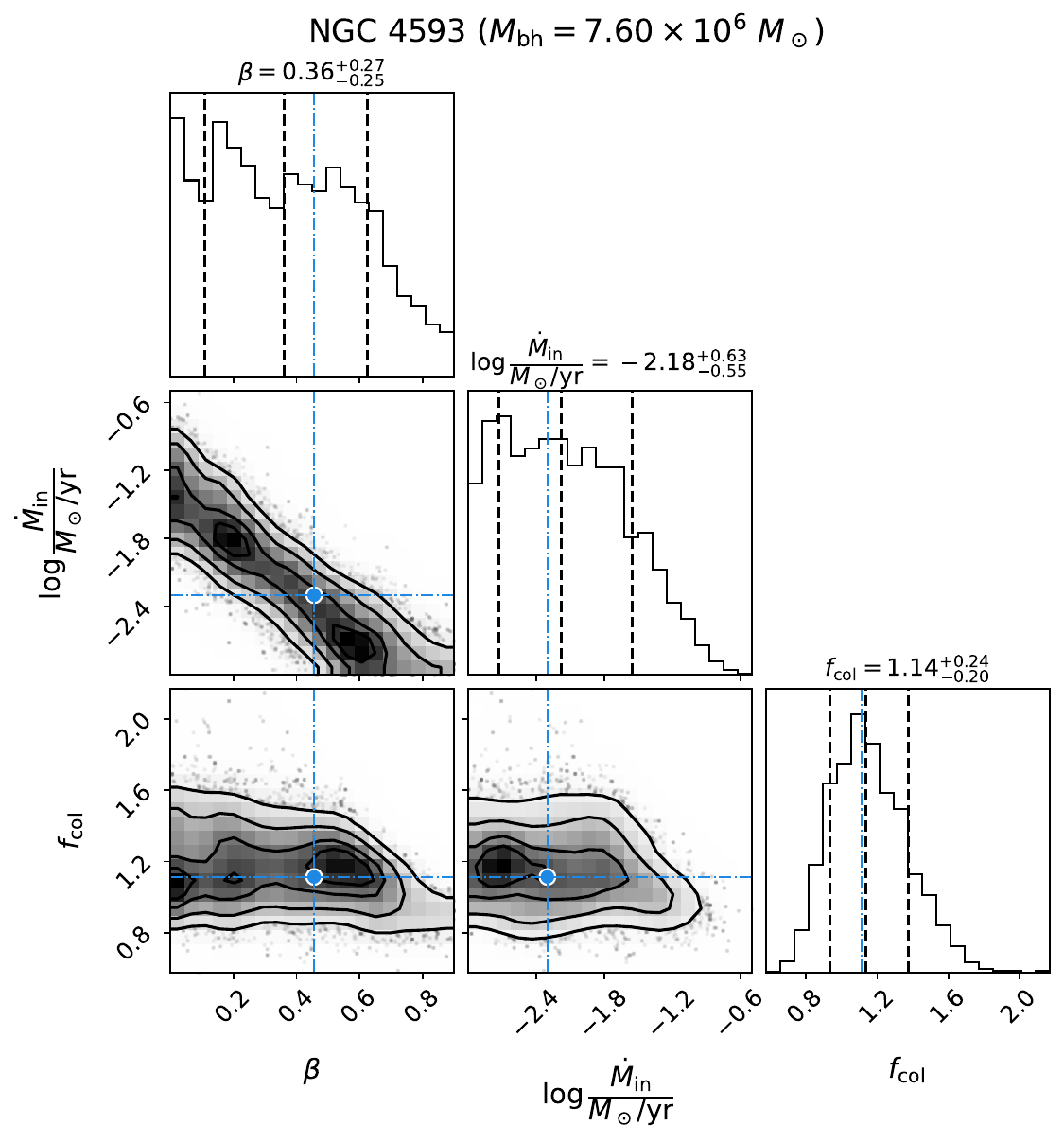}
    \includegraphics[width=0.48\textwidth]{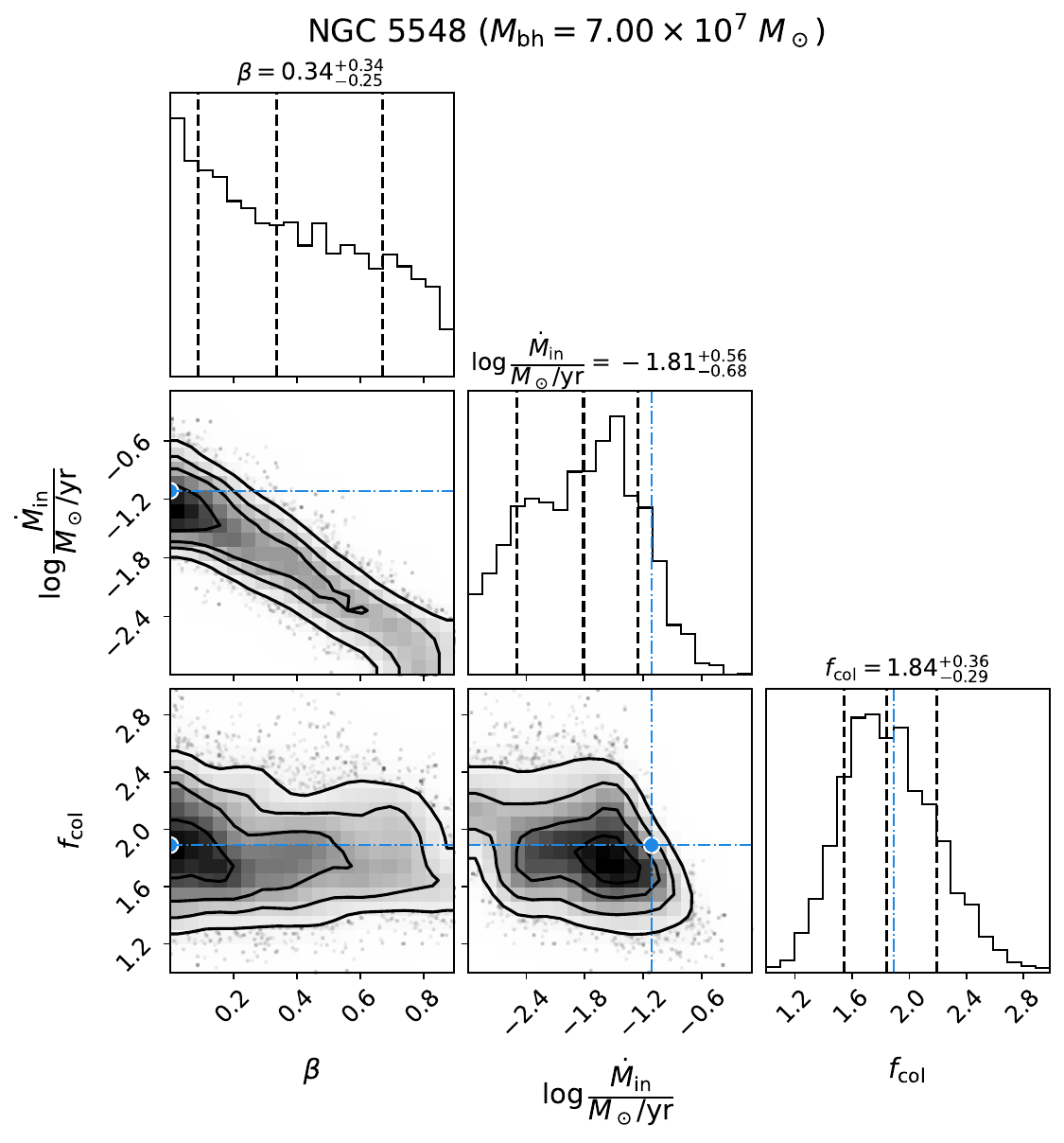}
    \includegraphics[width=0.48\textwidth]{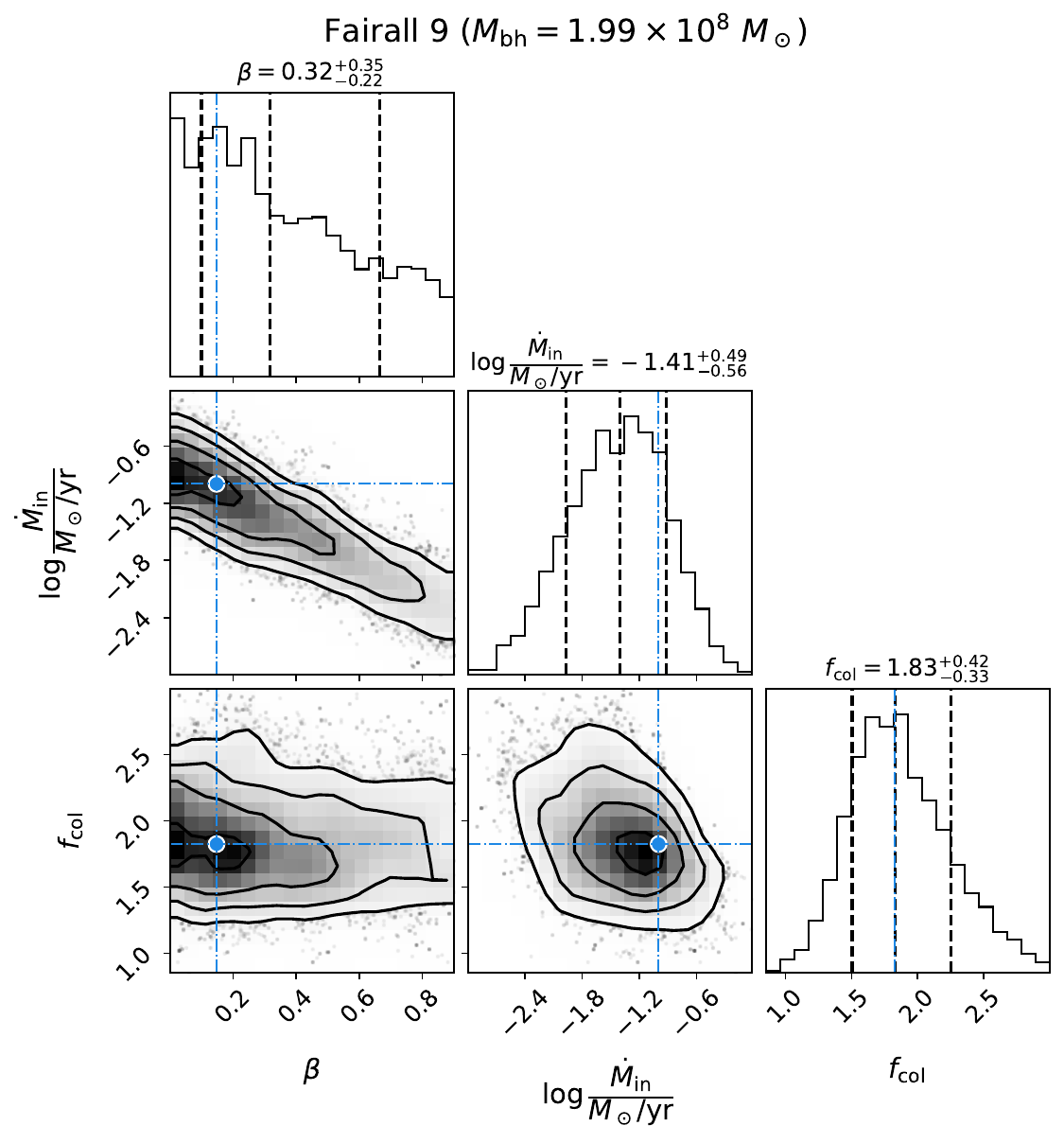}
    \includegraphics[width=0.48\textwidth]{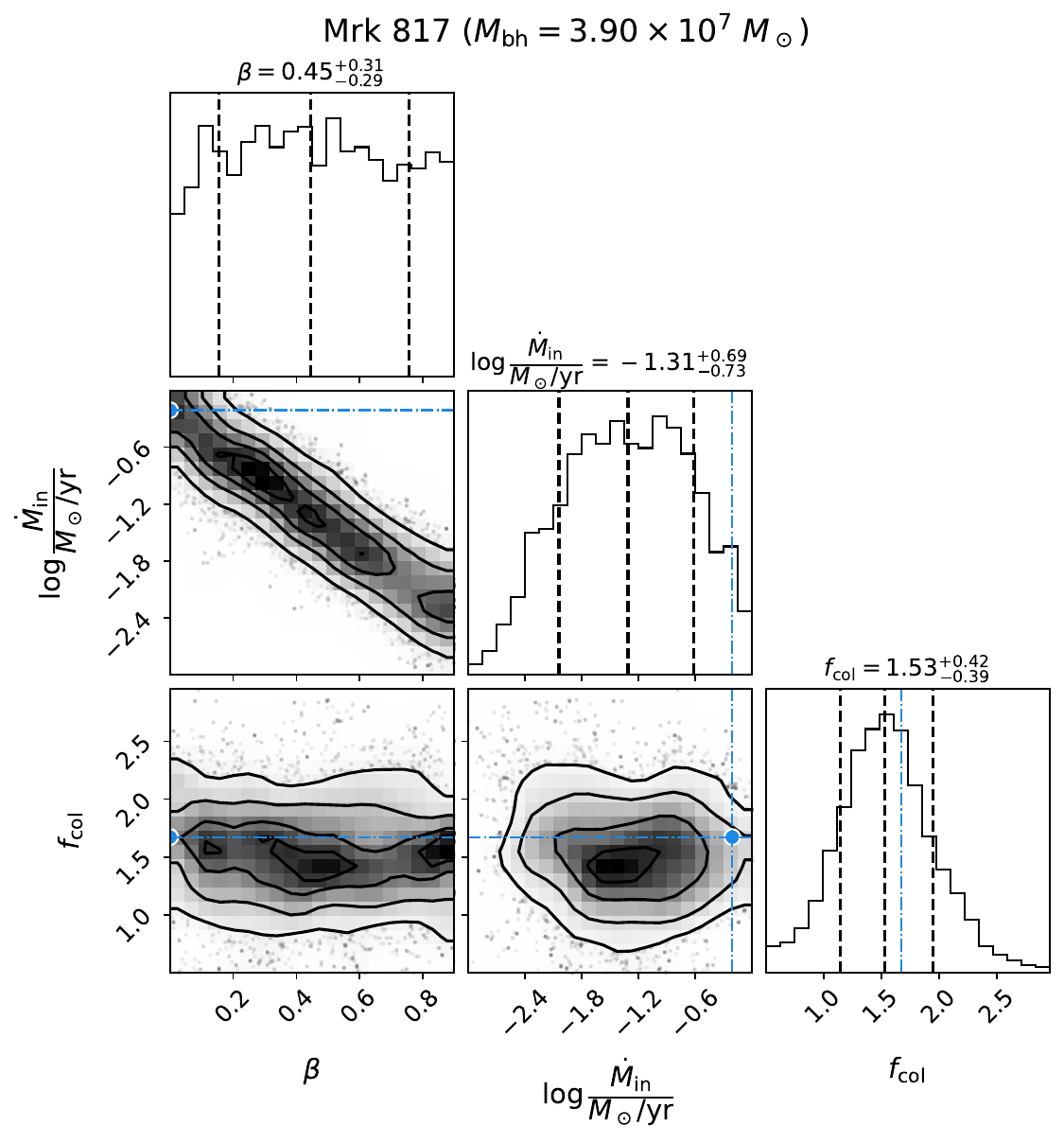}
    \caption{
    Sampling of MCMC.
    The contours enclose 68\%, 95\%, and 99.7\% of the sample.
    The vertical dashed black lines indicate the one standard deviation spread at the 16th, 50th, and 84th percentiles, with the corresponding numbers labeled on top of the 1-D distribution plot.
    Blue dots label the best-fit values from the Powell optimization, which may be trapped in local minima.
    }
    \label{fig:mcmc}
\end{figure}

Fitting the time lag spectra while allowing all parameters, including $\Mbh$, $a$, $\dMdt$, $\beta$, $i$, $H$, $\etaX$, $\fcol$, and $\Rout$, to vary freely is not feasible, as some parameters are inherently degenerate.
To address this, we propose an alternative approach for fitting the observed time lags and luminosity at $5100~\textrm{\AA}$ for the four AGNs in our sample.
First, we keep the spin $a=0$ (i.e. $\alpha=6$) since $\alpha$ is degenerate with $\dMdt$ (i.e. ${R_c}^{3-\beta}\propto {\nu L_\nu}^{\frac{3-\beta}{2}} \propto\dMdt\alpha^{-\beta}$).
Second, we fix the inclination to $i=0$ for this analysis, given that the test cases are type I AGNs; further discussion of this parameter is provided in \sref{sec:discussion}.
Last, we adopt the time lag derived from the power-law approximation in \eref{eqn:time_lag_pl}, which assumes $\Rin,H \ll R \ll \Rout$.
In essence, we reduce the parameters to three by relating them to one another, within the fitting ranges of $0\leq\beta<1$, $-3\leq\log\frac{\dMdt}{\Msun/{\rm yr}}<0$, and $0.5\leq\fcol<3$. 
In addition, we provide the fits with fixed $\fcol=1$ and with fixed $\beta=0$ for comparison, if needed.

The chi-square value for the time lag is defined as:
\begin{equation}
    \chitau = \frac{1}{N_\lambda}\sum_\lambda \frac{\left[\Delta\tau^{\rm meas}_{\lambda}-\left(\tau_\lambda-\tau_{\lambda_0}\right)\right]^2}{{\sigma_\tau}^2},
    \label{eqn:chitau}
\end{equation}
where $\Delta\tau_\lambda^{\rm meas}$ represents the measured time lag in the rest frame relative to the reference wavelength $\lambda_0 =\lambda_{\rm ref}/(1+z)$, $\sigma_\tau$ is the average of the upper and lower limits of the error bars, and $N_\lambda$ is the number of time lag measurements.
During the fitting process, we exclude measurements that may be influenced by emission from the much larger Broad Line Region (BLR) relative to the disk (see also \sref{sec:discussion} for further discussion).
We further adopt the measured luminosity at $5100~\textrm{\AA}$ as our prior, based on \eref{eqn:flux_pl}, defined as
\begin{equation}
    \chilum = \frac{\left(\log \nu L^{\rm meas}_\nu -\log \nu L_\nu\right)^2}{{\sigma_{\log \nu L_\nu}}^2},
    \label{eqn:chilum}
\end{equation}
where $\sigma_{\log \nu L_\nu}$ is chosen as $\log(0.5)$, indicating that the measurement can be deviated by a factor of 2 due to inclination or contamination of host galaxy.
We choose monochramatic luminosity instead of bolometric luminosity as our prior because the bolometric luminosity from literature is estimated considering $\beta=0$, which does not apply to this work.
The total chi-square is then defined as $\chi^2 = \chitau + \chilum$.

\begin{figure*}[t]
\centering
\includegraphics[width=\textwidth]{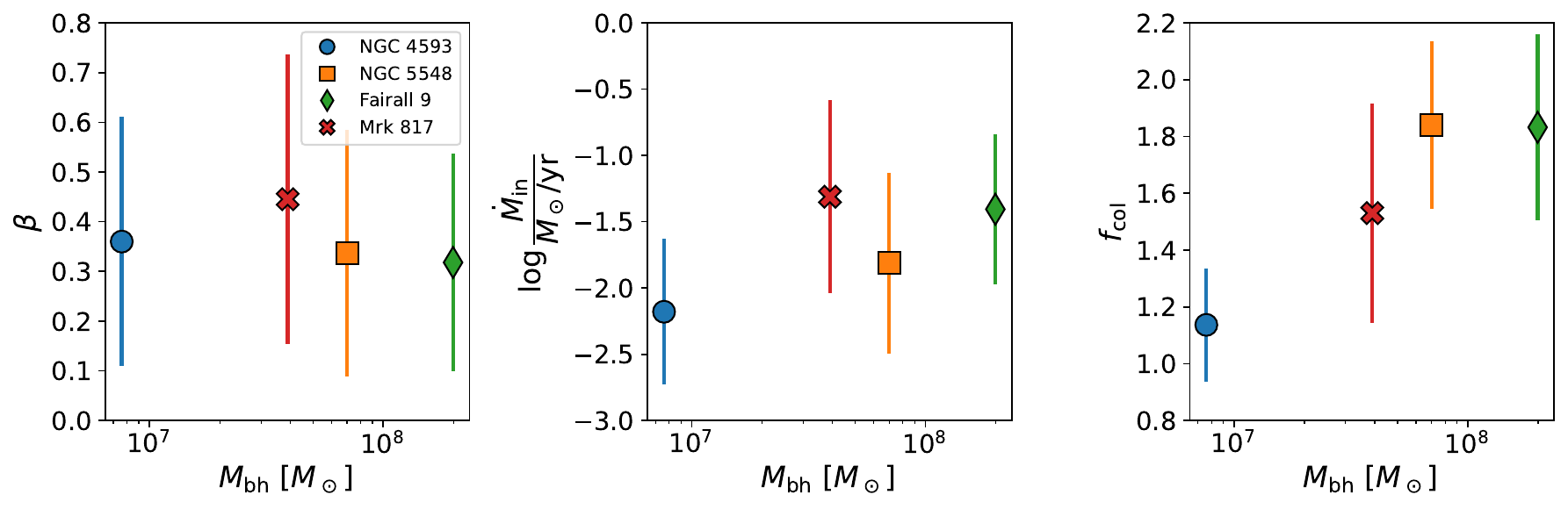}
\caption{
Relations of black hole mass $\Mbh$ with disk parameters: (\textit{left}) power index of the wind $\beta$, (\textit{middle}) accretion rate $\dMdt$, and (\textit{right}) color correction $\fcol$.
The disk parameters are obtained from the MCMC samples in \fref{fig:mcmc}, with the error bars indicating the 1-$\sigma$ uncertainty.
}
\label{fig:relation}
\end{figure*}
We observe that the fitting process can occasionally become trapped in local minima.
Therefore, alongside with the minimization of modified Powell algorithm\footnote{\url{https://docs.scipy.org/doc/scipy/reference/optimize.minimize-powell.html}}, we also employ Markov chain Monte Carlo (MCMC)\footnote{\url{https://emcee.readthedocs.io/en/stable/}} to investigate the correlation between fitting parameters.
The results are shown in \fref{fig:case}, while the MCMC samples are illustrated in \fref{fig:mcmc}.
Given the current data quality, breaking the degeneracy between $\beta$ and $\dMdt$ remains challenging.
We also observe that fitting based solely on $\chitau$ introduces a degeneracy between $\dMdt$ and $\fcol$ (i.e., ${R_c}^{3-\beta}\propto \dMdt {\fcol}^4$). 
This suggests that incorporating the flux spectrum or the spectral energy distribution (SED) into the fitting process could potentially offer a more robust solution, which may help break the degeneracy and provide a more accurate fit.
There are delays that tend to be smaller than the predicted values from the model at longer wavelengths ($\geq8000~{\rm \AA}$), particularly in the cases of NGC~4593 and NGC~5548, although the uncertainties are significant.
This discrepancy could be due to the outer edge of the disk, as the power-law approximation assumes $\Rout \to \infty$. 
One potential solution is to fit the data using a direct integral in \eref{eqn:time_lag}, which may slow down the fitting process.

Finally, we present the relationship between black hole mass and the fitting parameters of MCMC samples from our four AGNs in \fref{fig:relation}. 
Given the small sample size, we recognize the need for caution in making strong claims; however, this work still sheds light on certain aspects and sets the stage for future research with larger samples.
In the left panel, our MCMC samples indicate a shallower temperature slope, $\beta\approx0.35$, corresponding to a temperature slope of $T\propto R^{-0.66}$, which has also been reported in microlensing analyses \citep{BateEtal18,PoindexterEtal08,Cornachione&Morgan20}. 
However, the significant uncertainty in $\beta$ complicates the identification of any clear trend.
From the Powell best-fit (see \fref{fig:mcmc}), only one AGN (NGC~4593) shows a relatively higher $\beta = 0.46$, while the remaining AGNs can still be fitted by the traditional thin disk model.
From the middle panel, there is modest positive relation between black hole mass and accretion rate (except for Mrk~817), as also reported in Figure 8 in \cite{ShenEtal19}.
We note that their result is presented in $\Lbol$; it can be converted to accretion rate, though a more detailed examination of this conversion is necessary.
In the right panel, our best-fit yields $\fcol\approx1.6$, aligning well with simulation findings from \cite{Davis&El-Abd19}. 
Based on our results, we do not observe a trend in $\beta$, but we find a positive correlation between the accretion rate and color correction with black hole mass. 
We restate that given the small sample size and not insignificant uncertainties, it is difficult to assert any strong trends.

\section{Discussion}
\label{sec:discussion}
\begin{figure*}[t]
\centering
\includegraphics[width=\textwidth]{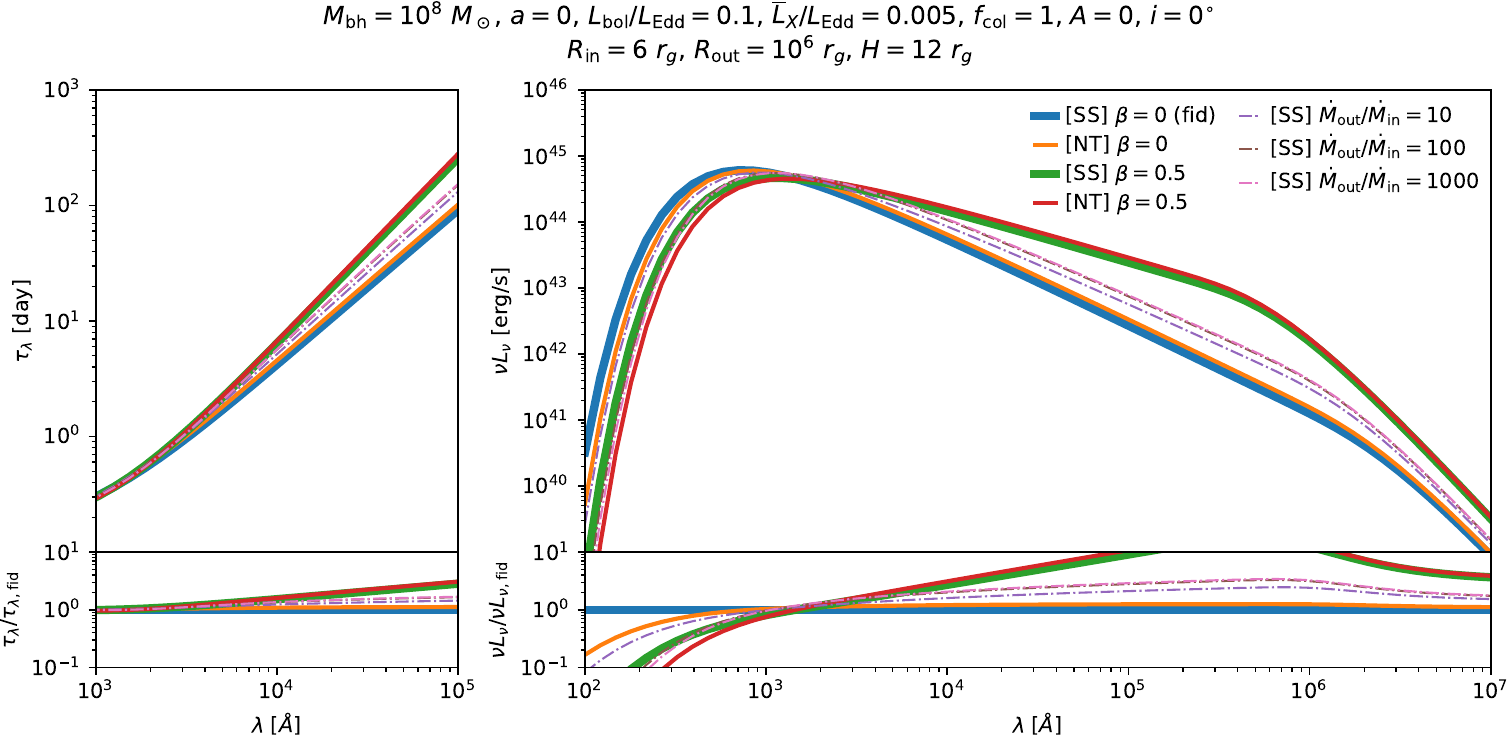}
\caption{
Spectra of average time lag (\textit{left}) and luminosity (\textit{right}) with varying disk and wind models.
The thick and thin solid lines represent the power-law accretion rate in \eref{eqn:dMdt} with the SS-disk and NT-disk, respectively, while the dash-dotted lines indicate the wind model in \eref{eqn:dMdt_BP82}.
The bottom panels illustrate the comparison between various cases and the fiducial one (the blue solid line labeled as ``fid''). 
The accretion disk parameters are indicated in the title. 
Note that the bolometric luminosity is constant across all lines.
}
\label{fig:mod}
\end{figure*}
In this section, we discuss the impact of alternative disk and wind models, examine the impact of the corona height and color correction, and cover other disk parameters, such as inclination and X-ray luminosity.
We also discuss about the continuum diffuse emission on reverberation mapping, and last we compare our simulation with the fitting formulas from the simulation of \cite{KammounEtal21a,KammounEtal23}.

\vspace{8pt}
\noindent\textit{Disk model}: 
We adopt the SS disk for our analysis, but replacing it with different disk models in the simulation is straightforward. 
An alternative approach is proposed by \cite{Novikov&Thorne73}, using a relativistic model for an optically thick, geometrically thin disk (referred to as the NT disk).
However, for UV/optical reverberation mapping, the SS-disk and NT-disk are indistinguishable, as demonstrated in \fref{fig:mod}.
It is important to note that the radiative efficiency for the NT disk is given by:
\begin{equation}
    \eta_{\rm NT}\equiv\sqrt{1-\left(1-\dfrac{2}{3\alpha}\right)},
\end{equation} 
to ensure a consistent bolometric luminosity \citep[e.g.][and references therein]{Bambi18}.

\vspace{8pt}
\noindent\textit{Wind model}: 
Although we primarily use the power-law accretion rate in this work, our simulations could also incorporate several wind models \citep[e.g.][]{YouEtal16}. 
One of the advantages with the power-law accretion rate is that it allows us to apply a power-law approximation for the temperature slope.
Here, we test a more physical wind model that describes magnetically driven winds \cite{Blandford&Payne82}.
The accretion rate is expressed as:
\begin{equation}
    \dot{M}(R)=\dot{M}_{\rm in} + \left(\dot{M}_{\rm out}-\dot{M}_{\rm in}\right)\frac{\ln(R/R_{\rm in})}{\ln(R_{\rm out}/R_{\rm in})},
    \label{eqn:dMdt_BP82}
\end{equation}
where $\dot{M}_{\rm in}$ and $\dot{M}_{\rm out}$ are free parameters, defined as accretion rates at $\Rin$ and $\Rout$, respectively.
The results, shown as dash-dotted lines in \fref{fig:mod}, indicate that this model can mildly increase the slope of the time lag spectrum, although it is not as effective as the power-law accretion rate. 
This is evident from comparing the extreme case of $\dot{M}_{\rm out}/\dot{M}_{\rm in} = 1000$ (dash-dotted pink line) with the case where $\beta = 0.5$, which yields $\dot{M}_{\rm out}/\dot{M}_{\rm in} = \left(\Rout/\Rin\right)^\beta \approx 400$ (when $\Rin = 6~r_g$ and $\Rout = 10^6~r_g$).
Note that the bolometric luminosity is fixed in this test.

\begin{figure*}
\centering
\includegraphics[width=\textwidth]{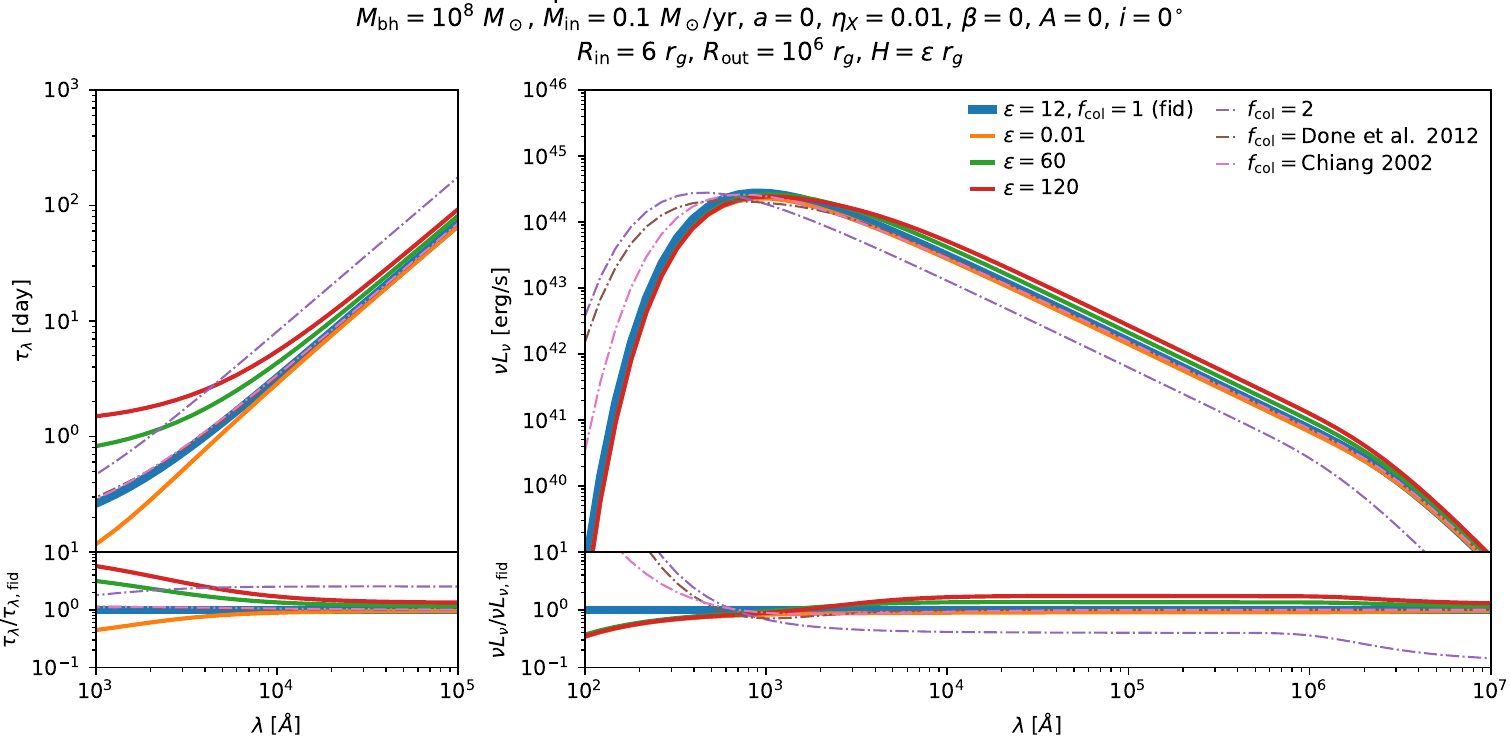}
\caption{
Spectra of average time lag (\textit{left}) and luminosity (\textit{right}) with varying  corona height $H$ and color correction $\fcol$.
The bottom panels illustrate the comparison between various cases and the fiducial one (the blue solid line labeled as ``fid''). 
The accretion disk parameters are indicated in the title. 
}
\label{fig:others}
\end{figure*}

\vspace{8pt}
\noindent\textit{Corona height}: 
The height of corona can enhance the time lag, particularly for the low wavelengths, mimicking shallower slope of time lag spectrum \citep{KammounEtal19}.
This is consistent with our findings, as shown in the right panel of \fref{fig:others}.
However, it has been reported that the corona likely resides within a few to tens of $r_g$ \citep[e.g.][]{HancockEtal23}, where the power-law temperature approximation remains valid.
The enhancement of size due to the corona height has also been applied to microlensing analysis \citep{PapadakisEtal22}, resulting in an increase in the half-light radius by a factor of $\approx3.5$. Nevertheless, it appears that they adopted a significantly higher X-ray luminosity for lensed quasars, with $\LX/\LEdd\approx0.1$, compared to nearby AGNs with $L_X/\LEdd\approx0.005$ (see further discussion below).
This suggests that the corona height has a relatively minor effect on reverberation mapping.

\vspace{8pt}
\noindent\textit{Color correction}: 
The impact of color correction is to enhance the disk size ($R_c\propto {\fcol}^{\frac{4}{3-\beta}}$) while keeping the bolometric luminosity unchanged.
This can mitigate the larger size with smaller Eddington ratio \citep[e.g. $\fcol=2.4$ adopted in][]{KammounEtal19,KammounEtal21b}.
We also observe this parameter is necessary in order to capture the data (see the dashed green lines in \fref{fig:case}).
Here, we test two additional models of color correction with function of effective temperature \citep[see Figure 1 in][]{ZdziarskiEtal22}: numerical fit in \cite{Chiang02} and observational fit in \cite{DoneEtal12}.
The results are illustrated as the dash-dotted lines in \fref{fig:others}.
We note that these two models do not affect the time lag spectra and luminosity at $5100~\textrm{\AA}$ significantly.

\vspace{8pt}
\noindent\textit{X-ray luminosity}:
The X-ray fluxes in the Seyfert 1 AGNs are relatively low, with $\LX/\LEdd\approx0.005$ \citep[see Table 1 in][]{UrsiniEtal20}, and should not affect the approximation of small temperature fluctuations.
However, the X-ray corona may not be the only lamp post; other sources of variability may also contribute to an increase in irradiation temperature, yielding larger amplitude of variability (see more discussion later in this section).
This is unlikely to significantly affect the data fitting of the power-law approximation, as the slope of the viscous temperature $\Tvisc \propto R^{(\beta-3)/4}$ tends to be shallower than that of the irradiation temperature $\Tlamp \propto R^{-3/4}$.
In the case of $\beta = 0$, the effect of irradiation can be absorbed into the viscous temperature, thereby reducing their impact on the fitting process.
An alternative is to fit the data using a direct integral, although it may slow down the fitting and introduce an excessive number of parameters.
We restate that the validity of the power-law approximation is based on the assumption that $\Tlamp < \Tvisc$.

\vspace{8pt}
\noindent\textit{Inclination angle}: 
The transfer function becomes more skewed at higher inclinations \citep[see Figure 1 in][]{StarkeyEtal17}. 
This increased skewness can also slightly decrease the mean due to the term of $H\cos i$ in \eref{eqn:tau}.
However, under the power-law temperature approximation, where $\Rin,H\ll R\ll \Rout$, inclination has no impact on the mean time lag; therefore, \eref{eqn:time_lag_pl} remains valid for fitting time lag spectra.
Inclination can reduce observed flux. 
For our current sample, which consists of type I AGNs with small inclinations, the analysis in this work is not affected by inclination significantly. 
Nevertheless, in microlensing analysis, where the disk orientation can interact with micro-caustic patterns, inclination can become a more significant parameter \citep[][]{Poindexter&Kochanek10}.

\vspace{8pt}
\noindent\textit{Albedo}:
In the UV/optical regime, thermal emission from the reprocessing of higher-energy radiation typically dominates the reflected component, leading to an albedo $A$ often assumed to be zero. 
However, near X-ray sources, we observe higher albedos, ranging from approximately 0.1 to 0.2 for hot, ionized accretion disks \citep{Haardt&Maraschi93,Liu&Qiao22}. 
A higher albedo can reduce irradiation heating, thereby aligning more closely with the power-law temperature approximation. 
While a more realistic albedo profile has been proposed \citep{Kazanas&Nayakshin01}, it does not affect the results presented in this work.

\vspace{8pt}
\noindent\textit{Outer radius}: 
There is a drop in luminosity (and time lag) at longer wavelengths ($\lambda\gtrsim 10^5~\textrm{\AA}$), affected by the outer radius $\Rout$ \citep[see also Figure 22 in][]{KammounEtal21a}.
The outer radius of the accretion disk remains uncertain, but can be estimated where the disk's self-gravity dominates over the central gravity of the black hole \citep[see Figure 6 in][]{YouEtal12}.
Fitting the spectral shape of the accretion disk in the infrared/optical wavebands could provide additional constraints on $\Rout$.

\vspace{8pt}
\noindent\textit{Radiative efficiency}: 
Although radiative efficiency ($\eta=\Lbol/\dot{M}c^2$) is not a free parameter in our model, it provides insight into how effectively an accretion disk converts the gravitational energy of infalling matter into radiation.
Estimating $\eta$ under a power-law accretion rate is challenging because the ratio $\dot{M}_{\rm out}/\dot{M}_{\rm in} = \left(\Rout/\Rin\right)^\beta$ can vary significantly. 
Typically, $\eta$ is calculated using $\dot{M}_{\rm out}$, which can result in a radiative inefficiency, particularly when $\beta>0.5$, leading to $\eta\ll0.1$ due to substantial outflows.
One way to address this issue is by introducing the spherization radius \citep{Shakura&Sunyaev73}, beyond which the accretion rate is restricted and cannot exceed the Eddington limit. 
However, this approach is relevant in the context of super-Eddington accretion flows. 
While this discussion is important, it falls outside the scope of this work.

\begin{figure}[t]
\centering
\includegraphics[width=\textwidth]{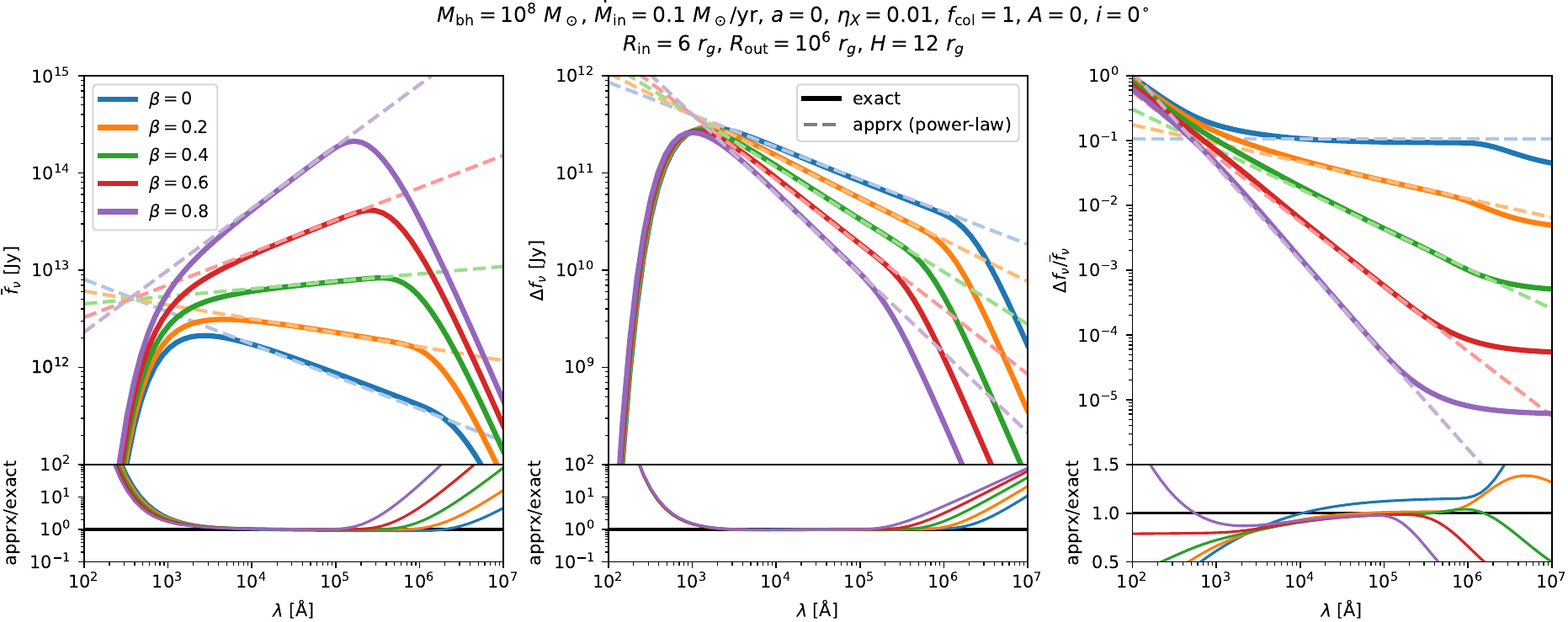}
\caption{
Spectra of flux (\textit{left}), amplitude of variation (\textit{middle}), and fractional variability amplitude (\textit{right}).
Given the power law approximation (shown as the dashed lines), the scaling relations are $\fmean \propto \lambda^{-\frac{1-3\beta}{3-\beta}}$, $\Delta f_\nu \propto\lambda^{-\frac{1+\beta}{3-\beta}}$, and $\Delta f_\nu/\fmean \propto\lambda^{-\frac{4\beta}{3-\beta}}$.
The accretion disk parameters are listed in the title, same as \fref{fig:lamb}.
}
\label{fig:flux}
\end{figure}
\vspace{8pt}
\noindent\textit{Amplitude of variability}:
This aspect is often overlooked in the reverberation mapping, as it focuses on temporal correlations and the structural response of the disk. 
In this work, we provide the estimates using the power-law approximation (\esref{eqn:flux_pl} and (\ref{eqn:dflux_pl})), deriving scaling relations for mean flux $\fmean \propto \lambda^{-\frac{1-3\beta}{3-\beta}}$, variability amplitude $\Delta f_\nu \propto\lambda^{-\frac{1+\beta}{3-\beta}}$, and fractional variability amplitude $\Delta f_\nu/\fmean \propto\lambda^{-\frac{4\beta}{3-\beta}}$, as illustrated in \fref{fig:flux}. 
According to the wind model, larger wavelengths correspond to smaller fractional variability amplitude, consistent with observations from the AGN sample (commonly denoted as $F_{\rm var}$ in the literature). 
However, matching these results to observed quantities requires considering the fluctuation of the driving source (i.e. $X'=X-1$), making direct comparison challenging.
We emphasize that to utilize the convolution approach or the power-law approximation, it is essential for $\Delta T \cdot X' < \Tmean$ to satisfy the Taylor expansion approximation, as described in \eref{eqn:temp_apprx}. 
Typically, the variability amplitude in X-rays is around $0.001$ - $0.1$ \citep{O'NeillEtal05}; even when $X'$ fluctuates on the order of 1, the power-law approximation remains valid as long as $\etaX\ll1$.
We examine the UV/optical fractional variability in our AGN sample ranging from $0.02$ to $0.2$, which can be described with the best-fit models (adopting $\etaX\approx0.03$).
While other sources may also influence observed variability, such as intrinsic disk fluctuations \citep[e.g.][]{HagenEtal24}, it lies beyond the scope of this work.

\vspace{8pt}
\noindent\textit{Diffuse continuum emission}:
Light from outer regions, such as the BLR, can contaminate the continuum emission and hence enhance the time lag \citep{CackettEtal18,Korista&Goad19,HernandezSantistebanEtal20,PozoNunezEtal23}. 
These discrepancies have been interpreted as contributions from the diffuse continuum emission (DCE) of the BLR, due to free–free and free–bound hydrogen transitions \citep{Korista&Goad01}. 
The imprint of the BLR in the lag spectrum is particularly reflected as longer lags towards the Balmer ($3646~\textrm{\AA}$) limit and the Paschen ($8204~\textrm{\AA}$) lines,  showing evidence for the Balmer edge and marginal evidence in the Paschen edge in emission.
In this work, we exclude time lag measurements potentially affected by emission from the BLR to ensure the accuracy of our results, specifically from $3000$ to $4000~\textrm{\AA}$.
Full consideration of the DCE lag spectrum will be explored in future work.

\begin{figure}[t]
\centering
\includegraphics[width=\textwidth]{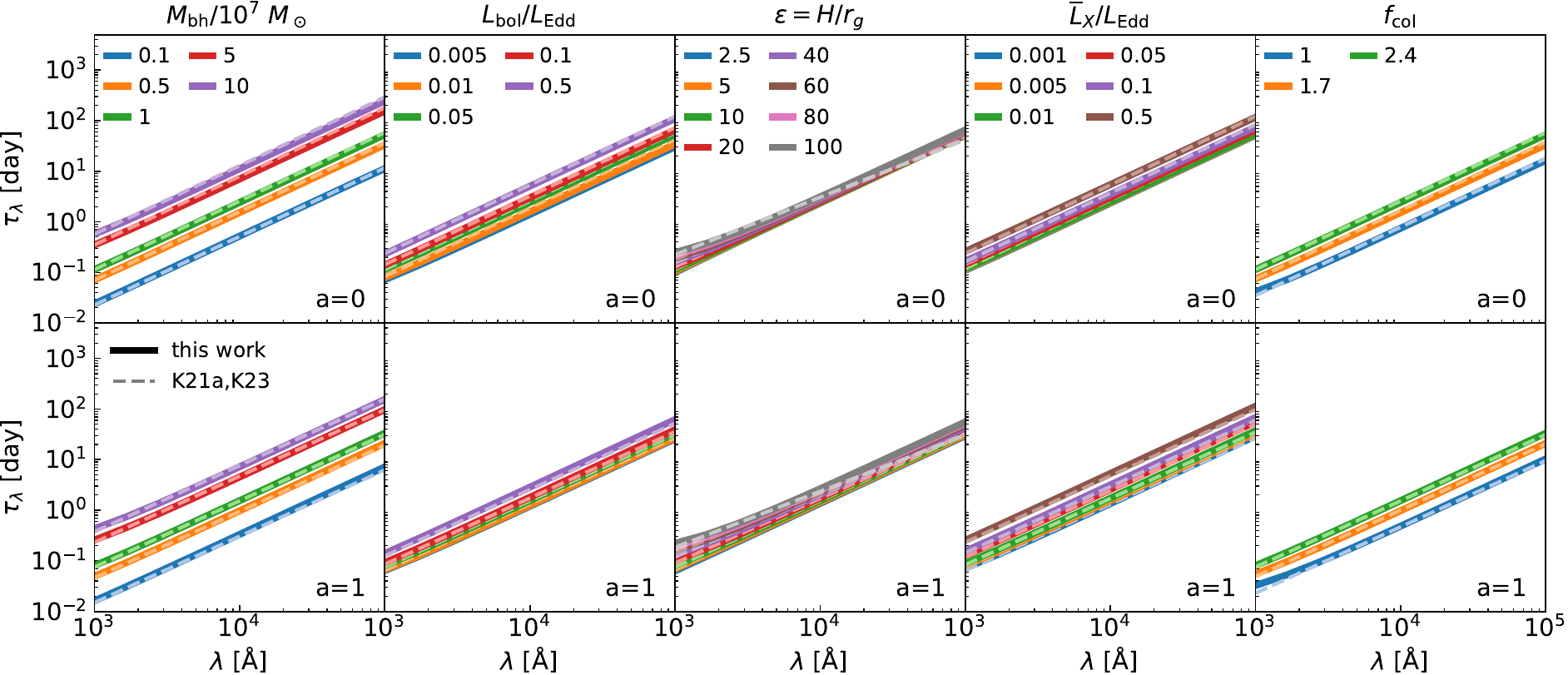}
\caption{
Comparison of time lags with fitting formulas, using reference parameters: $\Mbh=10^7~\Msun$, $\Lbol/\LEdd=0.05$, $\epsilon=10$, $\Lxmean/\LEdd=0.005$, $\fcol=2.4$, $\beta=0$, $i=0~\deg$, and $\Rout=10^6~\rg$. 
Each column shows the results with one parameter altered individually, with spin $a=0$ (\textit{top}) and $a=1$ (\textit{bottom}).
The solid lines represent our simulation results, while the dashed lines are from the fitting formulas of \cite{KammounEtal21a,KammounEtal23}, denoted as K21a and K23.
}
\label{fig:k21}
\end{figure}
\vspace{8pt}
\noindent\textit{Comparison with other simulations}:
Other models of disk reprocessing using general-relativistic ray-tracing have been developed in \cite{KammounEtal19}.
We adopt the fitting formulas from \cite{KammounEtal21a} and the updated ones including $\fcol$ from \cite{KammounEtal23}.
Their time-lag formulas has been applied to fit the UV/optical time-lags \citep{KammounEtal21b,KammounEtal23}.
In this test, we adopt the SS disk while they chose the NT disk, though the choices do not affect the result.
We compare most of parameters listed in Table 1 of \cite{KammounEtal21a}, with the result shown in \fref{fig:k21}.
This evaluation demonstrates that our model (without considering $\beta$) provides nearly identical relations.

\section{Conclusions}
\label{sec:conclusion}
We investigate the wavelength dependence of flux and time lags in AGN accretion disks, focusing on a geometrically thin and optically thick disk model irradiated by an external ``lamp post'' source. 
By introducing a power-law dependence for the accretion rate $\dot{M}=\dMdt\left(R/\Rin\right)^\beta$ inspired by disk wind dynamics \citep{Blandford&Begelman99}, we extend previous methodologies for reverberation mapping. 
Our analysis reveals that a higher $\beta$ results in a larger disk size and increased luminosity, offering an alternative approach to addressing the issue of accretion disk sizes.
Furthermore, we provide analytical formulas for more efficient application to observational data, particularly in the UV/optical regime.
Our fitting analysis of time lags and luminosity for type I AGNs, using a power-law approximation, reveals a shallower temperature slope ($T \propto R^{-0.66}$, or $\beta \approx 0.35$) compared to the traditional thin disk model ($T \propto R^{-3/4}$), and a color correction factor of $\fcol \approx 1.6$.
These results are consistent with previous studies \citep[e.g.][]{Davis&El-Abd19,Cornachione&Morgan20}.
However, due to high uncertainties, we cannot completely rule out the traditional thin disk model.
We observe a positive correlation between the accretion rate and color correction with black hole mass, though the small sample size limits definitive conclusions.
We also find a strong degeneracy between $\beta$ and accretion rate $\dMdt$, which complicates the fitting process. 
This suggests that incorporating the flux spectrum or spectral energy distribution into the fitting process could help break this degeneracy and provide a more accurate solution.
Lastly, we address various aspects including alternative disk and wind models, as well as other disk parameters, demonstrating that our simulation offers a fast and accurate approach for reverberation mapping.

Our work lays the foundation for more realistic simulations, enabling the rapid generation of quasar light curves that are applicable to diverse observational scenarios in training sets, while integrating machine learning modeling through latent stochastic differential equations \citep[latent SDEs;][]{FaginEtal24b,FaginEtal24}. 
This capability is crucial for addressing challenges associated with traditional curve-shifting techniques. 
Moreover, our efficient approach opens opportunities to combine two dimensional flux variability with microlensing analysis \citep[e.g.][]{ChanEtal21,BestEtal24,BestEtal24b}, allowing for a deeper investigation into the inner structure of the disk.

\section*{Acknowledgments}
We thank D.~Sluse, R.~Soria, O.~Blaes, I.~Papadakis, R.~Daly, S.~Ford and B.~McKernan for useful discussion.
We also appreciate the constructive feedback provided by the anonymous referee.
Support was provided by Schmidt Sciences, LLC. for J.~H.-H.~Chan, J.~Fagin, H.~Best, and M.~J.~O'Dowd.

\software{
\texttt{Numpy}~\citep{numpy}, 
\texttt{SciPy}~\citep{scipy}, 
\texttt{Astropy}~\citep{astropy}, 
\texttt{Matplotlib}~\citep{matplotlib}, 
\texttt{emcee}~\citep{emcee},
\texttt{corner.py}~\citep{corner}.
}

\bibliography{sample631}{}
\bibliographystyle{aasjournal}

\end{CJK*}
\end{document}

%% file: table/case.tex
\begin{tabular}{ccrcccr}
\hline\hline
\multicolumn{1}{c}{Name} & \multicolumn{1}{c}{Redshift} & \multicolumn{1}{c}{$\Dlumi$} & \multicolumn{1}{c}{$\Mbh$} & \multicolumn{1}{c}{$\nu L_\nu(5100\textrm{\AA})$} & \multicolumn{1}{c}{$\lambda_{\rm ref}$} & \multicolumn{1}{c}{Reference} \\
\multicolumn{1}{c}{} & \multicolumn{1}{c}{$z$} & \multicolumn{1}{c}{[${\rm Mpc}$]} & \multicolumn{1}{c}{[$10^8~M_\odot$]} & \multicolumn{1}{c}{[$10^{43}~{\rm erg}/{\rm s}$]} & \multicolumn{1}{c}{[\AA]} & \multicolumn{1}{c}{} \\
\hline
  NGC 4593 &     0.0087 &      37.51 &       0.08 &       0.74 &       1928 &                    \cite{CackettEtal18} \\
  NGC 5548 &     0.0172 &      74.55 &       0.70 &       2.30 &       1367 &                  \cite{FausnaughEtal16} \\
 Fairall 9 &     0.0470 &     208.59 &       1.99 &       7.90 &       1928 &       \cite{HernandezSantistebanEtal20} \\
   Mrk 817 &     0.0315 &     137.98 &       0.39 &       6.90 &       1928 &                       \cite{KaraEtal21} \\
\hline
\end{tabular}